\documentclass[longauth]{aa}
\usepackage{natbib}
\setcitestyle{round}
\usepackage{caption}
\usepackage{subcaption}
\usepackage{lipsum}
\usepackage{lscape}
\usepackage{amsmath}
\usepackage{xcolor}
\usepackage{booktabs}
\usepackage[symbol]{footmisc}
\usepackage{academicons}
\usepackage{orcidlink} 
\usepackage{bm}
\usepackage{array}
\usepackage{comment}

\newcolumntype{F}[1]{%
    >{\raggedright\arraybackslash\hspace{0pt}}p{#1}}%
\newcolumntype{T}[1]{%
    >{\centering\arraybackslash\hspace{0pt}}p{#1}}%

\usepackage{hyperref}
\hypersetup{
  colorlinks   = true, 
  urlcolor     = blue, 
  linkcolor    = blue, 
  citecolor   = blue 
}

\usepackage{graphicx}
\usepackage{txfonts}
\begin{document} 

   \title{Missing components in $\Lambda$CDM from DESI Y1 baryonic acoustic oscillation measurements: Insights from redshift remapping}

   \author{E. Fernández-García\orcidlink{0009-0006-2125-9590}\thanks{e-mail: efdez$@$iaa.es}\inst{1}
          \and
        R. Wojtak\orcidlink{0000-0001-9666-3164}\inst{2}
        \and
          F. Prada\orcidlink{0000-0001-7145-8674}\inst{1}
          \and
          J. L. Cervantes-Cota\orcidlink{0000-0002-3057-6786}\inst{3}
          \and
          O.~Alves\orcidlink{0000-0002-7394-9466}\inst{4}
          \and
          G.~Valogiannis\orcidlink{0000-0003-0805-1470}\inst{5, 6}
          \and
          J.Aguilar\inst{7}
          \and
          S. Ahlen\orcidlink{0000-0001-6098-7247}\inst{8}
          \and
          S.~BenZvi\orcidlink{0000-0001-5537-4710}\inst{9}
          \and
          D.~Bianchi\orcidlink{0000-0001-9712-0006}\inst{10, 11}
          \and
          D. Brooks\inst{12}
          \and
          T.~Claybaugh\inst{7}
          \and
          A.~de la Macorra\orcidlink{0000-0002-1769-1640}\inst{13}
          \and 
          P.~Doel\inst{12}
          \and
          E.~Gaztañaga\inst{14, 15}
          \and
          S.~Gontcho A Gontcho\orcidlink{0000-0003-3142-233X}\inst{7}
          \and
          G.~Gutierrez\inst{16}
          \and
          K.~Honscheid\orcidlink{0000-0002-6550-2023}\inst{17, 18, 19}
          \and
          M.~Ishak\orcidlink{0000-0002-6024-466X}\inst{20}   
          \and
          S.~Juneau\orcidlink{0000-0002-0000-2394}\inst{21}
          \and
          D.~Kirkby\orcidlink{0000-0002-8828-5463}\inst{22}
          \and
          T.~Kisner\orcidlink{0000-0003-3510-7134}\inst{7}
          \and
          M.~Landriau\orcidlink{0000-0003-1838-8528}\inst{7}
          \and
          L.~Le~Guillou\orcidlink{0000-0001-7178-8868}\inst{23}
          \and
          M.~E.~Levi\orcidlink{0000-0003-1887-1018}\inst{7}   
          \and
          T.~S.~Li\orcidlink{0000-0002-9110-6163}\inst{24}
          \and
          M.~Manera\orcidlink{0000-0003-4962-8934}\inst{25, 26}
          \and
          A.~Meisner\orcidlink{0000-0002-1125-7384}\inst{27}
          \and
          R.~Miquel\inst{25, 28}
          \and
          J.~Moustakas\orcidlink{0000-0002-2733-4559}\inst{29}
          \and
          A.~Muñoz-Gutiérrez\inst{13}
          \and
          S.~Nadathur\orcidlink{0000-0001-9070-3102}\inst{30}
          \and
          W.~J.~Percival\orcidlink{0000-0002-0644-5727}\inst{31, 32, 33}
          \and
          I.~P\'erez-R\`afols\orcidlink{0000-0001-6979-0125}\inst{34}
          \and
          G.~Rossi\inst{35}
          \and
          E.~Sanchez\orcidlink{0000-0002-9646-8198}\inst{36}
          \and
          M.~Schubnell\inst{4}
          \and
          H.~Seo\orcidlink{0000-0002-6588-3508}\inst{37}
          \and
          D.~Sprayberry\inst{21}
          \and
          G.~Tarl\'{e}\orcidlink{0000-0003-1704-0781}\inst{5}
          \and
          B.~A.~Weaver\inst{25}
          \and
          H.~Zou\orcidlink{0000-0002-6684-3997}\inst{38}
          }

\institute{Instituto de Astrofisica de Andalucia (CSIC), E18008 Granada, Spain
\and
DARK, Niels Bohr Institute, University of Copenhagen, Jagtvej 128, DK-2200 Copenhagen, Denmark
\and
Carreterra México-Toluca S/N, La Marquesa, Ocoyoacac, Edo. de México C.P. 52750, México
\and
University of Michigan, 500 S. State Street, Ann Arbor, MI 48109, USA
\and
Department of Astronomy \& Astrophysics, University of Chicago, Chicago, IL, 60637, USA
\and
Kavli Institute for Cosmological Physics, Chicago, IL, 60637, USA  
\and
Lawrence Berkeley National Laboratory, 1 Cyclotron Road, Berkeley, CA 94720, USA
\and
Physics Dept., Boston University, 590 Commonwealth Avenue, Boston, MA 02215, USA
\and
Department of Physics \& Astronomy, University of Rochester, 206 Bausch and Lomb Hall, P.O. Box 270171, Rochester, NY 14627-0171, USA
\and
Dipartimento di Fisica ``Aldo Pontremoli'', Universit\`a degli Studi di Milano, Via Celoria 16, I-20133 Milano, Italy
\and
INAF-Osservatorio Astronomico di Brera, Via Brera 28, 20122 Milano, Italy
\and
Department of Physics \& Astronomy, University College London, Gower Street, London, WC1E 6BT, UK
\and           
Instituto de F\'{\i}sica, Universidad Nacional Aut\'{o}noma de M\'{e}xico, Circuito de la Investigaci\'{o}n Cient\'{\i}fica, Ciudad Universitaria, Cd. de M\'{e}xico C.~P.~04510, M\'{e}xico
\and
Institut d'Estudis Espacials de Catalunya (IEEC), c/ Esteve Terradas 1, Edifici RDIT, Campus PMT-UPC, 08860 Castelldefels, Spain
\and
Institute of Cosmology and Gravitation, University of Portsmouth, Dennis Sciama Building, Portsmouth, PO1 3FX, UK
\and
Fermi National Accelerator Laboratory, PO Box 500, Batavia, IL 60510, USA
\and
Department of Physics, The Ohio State University, 191 West Woodruff Avenue, Columbus, OH 43210, USA
\and
The Ohio State University, Columbus, 43210 OH, USA
\and             
Center for Cosmology and AstroParticle Physics, The Ohio State University, 191 West Woodruff Avenue, Columbus, OH 43210, USA
\and
Department of Physics, The University of Texas at Dallas, 800 W. Campbell Rd., Richardson, TX 75080, USA
\and
National Science Foundation's National Optical-Infrared Astronomy Research Laboratory
\and
Department of Physics and Astronomy, University of California, Irvine, 92697, USA
\and
Sorbonne Universit\'{e}, CNRS/IN2P3, Laboratoire de Physique Nucl\'{e}aire et de Hautes Energies (LPNHE), FR-75005 Paris, France
\and
Department of Astronomy $\&$ Astrophysics, University of Toronto, Toronto, ON M5S 3H4, Canada
\and       
Institut de F\'{i}sica d’Altes Energies (IFAE), The Barcelona Institute of Science and Technology, Edifici Cn, Campus UAB, 08193, Bellaterra (Barcelona), Spain
\and
Departament de F\'{i}sica, Serra H\'{u}nter, Universitat Aut\`{o}noma de Barcelona, 08193 Bellaterra (Barcelona), Spain
\and
NSF NOIRLab, 950 N. Cherry Ave., Tucson, AZ 85719, USA
\and
Instituci\'{o} Catalana de Recerca i Estudis Avan\c{c}ats, Passeig de Llu\'{\i}s Companys, 23, 08010 Barcelona, Spain
\and
Department of Physics and Astronomy, Siena College, 515 Loudon Road, Loudonville, NY 12211, USA
\and
Institute of Cosmology and Gravitation, University of Portsmouth, Dennis Sciama Building, Portsmouth, PO1 3FX, UK
\and
Department of Physics and Astronomy, University of Waterloo, 200 University Ave W, Waterloo, ON N2L 3G1, Canada
\and
Perimeter Institute for Theoretical Physics, 31 Caroline St. North, Waterloo, ON N2L 2Y5, Canada
\and
Waterloo Centre for Astrophysics, University of Waterloo, 200 University Ave W, Waterloo, ON N2L 3G1, Canada
\and
Departament de F\'isica, EEBE, Universitat Polit\`ecnica de Catalunya, c/Eduard Maristany 10, 08930 Barcelona, Spain
\and
Department of Physics and Astronomy, Sejong University, 209 Neungdong-ro, Gwangjin-gu, Seoul 05006, Republic of Korea
\and
CIEMAT, Avenida Complutense 40, E-28040 Madrid, Spain
\and
Department of Physics \& Astronomy, Ohio University, 139 University Terrace, Athens, OH 45701, USA
\and   
National Astronomical Observatories, Chinese Academy of Sciences, A20 Datun Rd., Chaoyang District, Beijing, 100012, P.R. China
}

   \date{Received ..., ...; accepted ..., ...}

\titlerunning{Missing information from the standard model}

 
  \abstract
   {}
   {We explore transformations of the Friedman-Lemaître-Robertson-Walker (FLRW) metric and cosmological parameters that align with observational data while aiming to gain insights into potential extensions of standard cosmological models.}
   {We modified the FLRW metric by introducing a scaling factor, $e^{2\Theta(a)}$--the cosmological scaling function (CSF), which alters the standard relationship between cosmological redshift and the cosmic scale factor without affecting angular measurements or cosmic microwave background (CMB) anisotropies. Using data from DESI Year 1, Pantheon+ supernovae, and the Planck CMB temperature power spectrum, we constrained both the CSF and cosmological parameters through a Markov chain Monte Carlo approach.
   }
   {Our results indicate that the CSF model fits observational data with a lower Hubble constant (although it is compatible with the value given by Planck 2018 within 1$\sigma$) and is predominantly dark matter dominated. Additionally, the CSF model produces temperature and lensing power spectra similar to those predicted by the standard model, though with lower values in the CSF model at large scales. We also checked that when fitting a CSF model without dark energy to the data, we obtain a more negative conformal function. This suggests that the CSF model may offer hints about missing elements and opens up a new avenue for exploring physical interpretations of cosmic acceleration.
   }
   {}


   \maketitle
%

\section{Introduction}

The Lambda cold dark matter ($\Lambda$CDM) cosmological model is the simplest model of the Universe mathematically, consisting of just two energetically dominant components at late times. Namely, these are the positive cosmological constant ($\Lambda$ > 0), which approximates dark energy, and CDM. This model is regarded as the most successful cosmological model introduced to date, as it is consistent with observational data from various astrophysical and cosmological probes, such as the accelerated expansion of the Universe \citep{1998AJ....116.1009R, 1999ApJ...517..565P}; the power spectrum and statistical properties of the cosmic microwave background (CMB; \citep{Page_2003}) and large-scale structures (LSSs) in the Universe; and hydrogen, deuterium, and helium abundances \citep{RevModPhys.88.015004}. In fact, the 2018 legacy release from the Planck satellite \citep{planck2018} of the CMB anisotropies provides strong support for the standard $\Lambda$CDM cosmological model. However, there are statistically significant tensions in the measurement of various quantities between the CMB data, which are cosmological model dependent and obtained assuming a $\Lambda$CDM model, and late-time cosmological model-independent probes.

The most statistically significant and long-standing tension is in the estimation of the Hubble constant, H$_{0}$, between the CMB data—which are dependent on the cosmological model and derived under the assumption of a $\Lambda$CDM model—and direct local distance ladder measurements. In particular, we refer to the Hubble tension as the 5.0$\sigma$ disagreement between the Planck collaboration \citep{planck2018}, which gives H$_{0}$ = (67.27 $\pm$ 0.60) km s$^{-1}$ Mpc$^{-1}$ at the 68$\%$ confidence level, and the latest 2021 SH0ES collaboration \citep{Riess_2022}, which provides H$_{0}$ = (73.04 $\pm$ 1.04) km s$^{-1}$ Mpc$^{-1}$ at the 68$\%$ confidence level, based on supernovae calibrated by Cepheids. However, it is important to note that there are not only these two values but actually two sets of measurements. All the indirect, model-dependent estimates at early times agree with each other, such as those from the CMB and baryonic acoustic oscillation (BAO) experiments. Similarly, the same agreement holds for all the direct late-time $\Lambda$CDM-independent measurements, such as those from distance ladders and strong lensing \citep[see][for more details]{Abdalla_2022}.
Nevertheless, several studies (e.g., \cite{2022MNRAS.515.2790W, 2024MNRAS.533.2319W}) have suggested that this tension in H$_{0}$ can be significantly reduced to 2.8$\sigma$ if alternative extinction models are applied to account for differing host galaxy properties in the calibration of Type Ia supernovae (SNe Ia), such as adjusting the total-to-selective extinction coefficient, $R_B$, and modifying the reddening distribution to better align with that observed in high-mass host galaxies.

A possible alternative to the standard model that could provide insights into what is missing involves consideration of conformal gravity, which was first introduced by \cite{WeylReineI}. These types of gravity theories lead to late-time accelerated expansion by modifying the Lagrangian of general relativity (GR). Conformal transformations of any metric can be expressed as

\begin{equation}
    \tilde{g}_{\mu\nu} = \Omega^{2}(x)g_{\mu\nu},
\end{equation}
where $\Omega(x)$ is the Weyl or conformal transformation. It is a non-vanishing regular function that affects the lengths of the time-space-like intervals and the norm of the time-space-like vectors, but it leaves the light-cones unchanged \citep{faraoni1998conformal}; that is, it  does not affect null geodesics and it preserves angles. Therefore, an advantage of these type of transformations is that they offer a compelling avenue for exploring substantial deviations from standard cosmological models without disturbing the observed smoothness of the CMB \citep{Visser_2015}.   

Considering these conformal transformations of the metric implies reconsidering the validity of the Friedman-Lemaître-Robertson-Walker (FLRW) metric. This has already been explored in previous works \cite[e.g.,][]{Kolb:2011zz, CLIFTON_2013, deledicque2020dark}, and the investigations in such studies are based on the fact that the homogeneity and isotropy of the Universe are not exact properties on large scales but approximate ones. Therefore, using the FLRW metric, which is an approximation of the Universe, can lead to cosmological parameter values that do not agree with the true ones. Hence, it is important to differentiate between the approximate metric ($g_{\mu\nu}$) and the true one ($\tilde{g}_{\mu\nu}$).

The interpretation of $\Omega(x)$ within this framework goes beyond a simple parametrization. Instead, it serves as a tool to capture any deviations between the true metric and the standard one. These deviations could manifest as a time-dependent evolution of cosmological parameters, including the cosmological constant (which may not remain constant over time), or as missing components in the standard model. However, the goal of our approach is not to impose a specific functional form on $\Omega(x)$ but to infer it directly from observational data in a non-parametric and semi-cosmographic manner. This involves consistently calculating all observables using the conformal FLRW metric while preserving the expansion history defined by the Hubble parameter, as given in standard FLRW cosmology. By maintaining the functional form of the Hubble parameter from GR, we ensure that any deviations from $\Omega \equiv 0.3$ arise from local effects or processes independent of gravity. This approach can also serve as a consistency test for GR, where recovering the standard GR elements confirms the validity of the framework, while deviations could either indicate limitations of the standard cosmological model or reflect inconsistencies in prior assumptions.

In this work, we fit BAO distances from DESI Year 1 (DESI Y1; \citep{desicollaboration2024desi}), the Pantheon+ supernova sample \citep{2022ApJ...938..110B}, and the Planck 2018 temperature power spectrum \citep{planck2018} to this conformal model, and we constrain the new cosmological parameters. We demonstrate that this new model is compatible with the CMB temperature, matter, and lensing potential power spectra predicted by the standard $\Lambda$CDM model and that we can recover Planck's parameters \citep{planck2018} by requiring that the comoving radial distance given by both models is the same.

A similar approach has been investigated in \cite{bassett2015observational}, \cite{2016MNRAS.458.3331W}, and \cite{2017MNRAS.470.4493W}. However, the first two studies assumed an overly simplistic relationship between the observed redshift and that predicted by the FLRW metric. Moreover, they relied on a limited number of BAO measurements and used binned  SN Ia datasets, leading to significantly large uncertainties in their results. In contrast, \cite{2017MNRAS.470.4493W} considered a sufficiently general relationship between the observed and FLRW redshifts (a spline function, which we also use in this work) and combined BAO and SN Ia data with CMB for the first time for this type of model. However, \cite{2017MNRAS.470.4493W} focused exclusively on open CDM models, preventing a direct comparison with our results, as we examine flat $\Lambda$CDM and CDM models.

This work is structured as follows: In Section \ref{redshift_remapping}, we introduce the conformal FLRW metric that we use, the conditions under which the energy-momentum tensor is conserved in this new metric, and the impact this metric has on the relationship between the observed redshift and the scale factor. Next, in Section \ref{observables}, we describe how the equations governing key observables (mainly distances, the Hubble parameter, and CMB temperature) change with this new definition of observed redshift. In Section \ref{spline}, we explain two different methods for calculating the conformal transformation. The first compares it with other cosmological models, and the second involves constraining its shape using observational data. In Section \ref{data}, we present the observational data used to obtain the conformal transformation of the FLRW metric. Finally, in Section \ref{results}, we show the constrained values for all cosmological parameters as well as the parameters involving the conformal transformation for flat conformal $\Lambda$CDM models. We also demonstrate that the temperature, matter, and lensing potential power spectra predicted by this model are compatible with those predicted by the standard $\Lambda$CDM model. Lastly, we show that we can recover Planck's parameters \citep{planck2018} by requiring that the comoving radial distance given by both models is the same.

\section{Conformal Friedman-Lemaître-Robertson-Walker metric and the impact on observational redshift}\label{redshift_remapping}
\subsection{The conformal FLRW metric}
In this work, we use the conformally distorted FLRW metric proposed in \cite{Visser_2015}:

\begin{equation}
    ds^{2}=e^{2\Theta(a)}\left[c^{2}dt^{2}-a(t)^{2}\left(\frac{dr^{2}}{1-kr^{2}} +r^{2}d\Omega\right)\right],
    \label{remapping}
\end{equation}
where $k=\pm 1, 0$ describes the spatial curvature of the Universe,  $\Theta(a)$ is a free function of $a(t)$ that parametrizes a conformal transformation, and $t$ is the cosmic time as measured by clocks characterized by $\Theta=0$. 
Conformal transformations stretch time and spatial distances while preserving the angles and null geodesics in the standard FLRW metric. This property is particularly important for the interpretation of CMB radiation. As with the FLRW metric, the conformal FLRW metric automatically guarantees the isotropy of the CMB on large angular scales, and from this perspective, the new metric is the only plausible generalization of the FLRW metric that does not violate the Copernican principle.

It is important to note that a general conformal transformation depends on both time and spatial coordinates, $\Theta(t, \bar{x})$. However, we can decompose this function into a time-dependent component and a spatially dependent component, averaging the spatial part over large scales to $<\exp{(\Theta(\bar{x}))}> = 1$, as we are only interested in large scales in this work. Thus, we can effectively consider that the conformal function depends only on time, $\Theta(a)$, and derive its dependence on redshift. Hereafter, we will refer to this specific type of conformal function as the cosmological scaling function (CSF).

An important exercise to perform was to check if the energy-momentum tensor is conserved. In order to do this, we followed the same procedure as in \cite{2008arXiv0806.2683D}. Therefore, we considered that the imposition of the conservation law in the standard frame gives in the conformally related frame

\begin{equation} 
    \tilde{T}^{ab}_{m;b}=-\frac{\Omega'^{a}}{\Omega}\tilde{T}_{m},
\label{Tab_eq}
\end{equation}
where $\tilde{T}$ is the energy-momentum tensor in the conformal frame; $\Omega$ is the conformal factor, $\Omega=\exp{(2\Theta)}$ in our case; $\Omega'^{a}$ is the derivative of $\Omega$ with respect to $a$; and the symbol $;$ is the covariant derivate.

We also checked whether the evolution of the density parameters remains the same as in the FLRW metric solving the previous equation for the time component. In order to solve this equation, we had to take into account that 

\begin{equation}
    \frac{\Omega_{'0}}{\Omega}T_{m}=-2 \dot{a} \frac{\partial\Theta}{\partial a}(\rho-3p)=-2 \dot{a} \frac{\partial\Theta}{\partial a}(1-3w)\rho
\end{equation}
and that

\begin{equation}
    \tilde{T}^{0b}_{m;b}=\dot{\rho}+3\frac{\dot{a}}{a}(1+w)\rho.
\end{equation}

By equating these two equations, we obtained

\begin{equation}
    \dot{\rho}+3\frac{\dot{a}}{a}\left(1+w + \frac{1}{3}\frac{a\,\partial\Theta}{\partial a}(1-3w) \right)\rho = 0.
    \label{rho_eq}
\end{equation}

One can see that this equation is the same as the one obtained in the FLRW metric but with an additional component, $\displaystyle{A=2a(\partial\Theta/\partial a)(1-3w)}$. Therefore, if one wants to maintain the same evolution for the density parameters as in the FLRW model, the term $\displaystyle{a,\partial\Theta/\partial a}$ must be negligible. This implies that $\Theta$ must be approximately constant, especially for high redshifts, as we show later.

At the decoupling epoch, $a \sim 10^{-3}$, and as highlighted in \cite{Visser_2015}, $\Delta[\Theta(a)_{\rm last scattering}] \leq 10^{-5}$. Therefore, one can easily see that one recovers the same evolution of the density parameters as in the FLRW metric.

For the late Universe, we expected $\Theta(a)$ to be sufficiently constant so that the evolution of dark energy and matter with time remains approximately the same as in the FLRW metric. In this way, the dynamics and structure formation in the CSF model remains the same as in the standard model, and the evolution of the density parameter with the scale factor would be 

\begin{equation}
    \rho(a) \sim \rho_{0}\exp{\left(3\int_{a}^{1}d\ln{(a)}(1+w(t))\right)},
\end{equation}
 as in the FLRW metric.

\subsection{Impact on observational redshift}
The main implication of introducing a CSF in the FLRW metric is breaking the standard relation between observed cosmological redshift and cosmic scale factor, i.e
\begin{equation}
    1+z_{\rm FLRW}=1/a.
\end{equation}

For the metric given by equation \ref{remapping}, cosmological redshift results both from the expansion of space and the evolution of $\Theta(a)$ and is given by

\begin{equation}
    1+z_{\rm obs} = \frac{a_{0}}{a_{\rm em}}\frac{e^{\Theta_{0}}}{e^{\Theta_{\rm em}}},
\end{equation}
where the subscripts mark the moment of emission (em) and observation (0). Exact values of $a$ and $\Theta$ at the present time have no fundamental physical meaning, and they merely set the units of our local measurements and observations. Therefore, without loss of generality, hereafter we use $a_{0}=1$ and $\Theta(a=1)=\Theta_{0}=0$ and drop subscripts (em), i.e.

\begin{equation}
    1+z_{\rm obs} = \frac{1}{ae^{\Theta}},
    \label{eq8}
\end{equation}
where $a = 1/(1+z_{\rm FLRW})$. Therefore, 

\begin{equation}
    \Theta(z_{\rm obs}) = -\ln\left[\frac{1+z_{\rm obs}}{1+z_{\rm FLRW}}\right],
    \label{thetalog}
\end{equation}
so we needed to relate $z_{\rm obs}$ to $z_{\rm FLRW}$ to obtain a function of $\Theta$ that only depends on this observed redshift or suppose a general parametrization for $\Theta(z_{\rm obs})$ and constrain these parameters with observational data.

In this framework, the relationship between the observed redshift and the standard FLRW redshift is modified due to the additional evolution introduced by $\Theta(a)$. This deviation can be interpreted as a remapping of redshifts \cite[see][for more details]{2017MNRAS.470.4493W, 2016MNRAS.458.3331W}, where the observed redshift $z_{\rm obs}$ no longer directly corresponds to the usual FLRW expansion history. Instead, it encodes both the standard cosmological redshift and the effects of $\Theta(a)$, effectively redefining the link between redshift and cosmic time. This "redshift remapping" provides a way to test deviations from the standard metric by analysing how distances inferred from cosmological observables are altered relative to the $\Lambda$CDM expectations.

\section{Observables}\label{observables}
In order to constrain the shape of $\Theta(z_{\rm obs})$ and all the involved cosmological parameters, we used observational constraints from BAO and SN Ia distances as well as the CMB power spectrum. In this section, we introduce the fundamental equations that involve all these quantities and explain how they must be rewritten with the CSF model considered in this work.

\subsection{Distances}
The most direct comoving distance we can define is the line-of-sight comoving distance:

\begin{equation}
    D_{C}(z_{\rm obs}) = c\int_{0}^{z'}\frac{dz_{\rm FLRW}}{H_{a}(z_{\rm FLRW})},
\label{DC}
\end{equation}
where
\begin{equation}
    1+z_{\rm obs}=(1+z')\frac{1}{e^{\Theta(z_{\rm obs})}}
\label{zprime}
\end{equation}
and
\begin{equation}
\begin{aligned}
    \frac{H_{a}^{2}}{H_{a0}^{2}} = \Omega_{\rm m}a^{-3}+\Omega_{\Lambda}\exp{\left(3\int_{a}^{1}\frac{1+w(a')}{a'}da'\right)},
    \label{Friedman_de}
\end{aligned}
\end{equation}
with $\displaystyle{H_{a}(a) = \dot{a}/{a}}$ being the Hubble parameter defined by the FLRW metric, i.e., the logarithmic expansion rate of the Universe. 

It is important to remark that the expression written for $H_{a}$ is an approximate one that takes into account the approximation that the evolution of the density parameters are the same as in the standard model, which was stated before. 

It can be seen that the only difference between equation \ref{DC} and the standard definition of $D_{C}(z_{\rm obs})$ when $z_{\rm obs}=z_{\rm FLRW}$ \citep{hogg2000distance} is the upper limit of the integral. Both definitions coincide when $\Theta(z_{\rm obs})=0$.

With these two distances, we could define the transverse comoving distance:
\begin{equation}
    D_{\rm M}(z_{\rm obs}) = \left\{ \begin{array}{lcc} 
        \displaystyle{D_{\rm H}(z_{\rm obs}) \frac{1}{\sqrt{\Omega_{k}}} \sinh \left( \sqrt{\Omega_{k}} \frac{D_{C}(z_{\rm obs})}{D_{\rm H}(z_{\rm obs})} \right)} & \text{if} & \Omega_{k} > 0 \\ 
        \\ 
        \displaystyle{D_{C}(z_{\rm obs})} & \text{if} & \Omega_{k} = 0 \\ 
        \\  
        \displaystyle{D_{\rm H}(z_{\rm obs}) \frac{1}{\sqrt{|\Omega_{k}|}} \sin \left( \sqrt{|\Omega_{k}|} \frac{D_{C}(z_{\rm obs})}{D_{\rm H}(z_{\rm obs})} \right)}  & \text{if} & \Omega_{k} < 0 
    \end{array} \right..
\end{equation}

This transverse comoving distance is related with the angular distance in the following way:  

\begin{equation}
    D_{A}(z_{\rm obs})=\frac{D_{\rm M}(z_{\rm obs})}{1+z_{\rm obs}}.
\end{equation}

As can be seen, these two distances, ($D_{\rm M}(z_{\rm obs})$ and $D_{A}(z_{\rm obs})$, depend on $\Theta(z_{\rm obs})$ only through $D_{c}(z_{\rm obs})$.

Apart from distances, BAO observations allow us to determine the Hubble constant and measure the Hubble parameter at different redshifts. In order to calculate the observed Hubble constant, we can rewrite equation \ref{DC} in the regime of small redshifts, i.e.

\begin{equation}
\begin{aligned}
    D_{C}(z_{\rm obs}) &= \frac{cz_{\rm obs}}{H_{a}(z_{\rm FLRW}=0)}\frac{dz_{\rm FLRW}}{dz_{\rm obs}}(z_{\rm FLRW}=0)=\\&=\frac{cz_{\rm obs}}{H_{a0}}\frac{dz_{\rm FLRW}}{dz_{\rm obs}}(z_{\rm FLRW}=0).
    \label{Dcz0}
\end{aligned}
\end{equation}

One can see now that the observed Hubble constant, H$_{0}$, is a net effect of the space expansion and the evolution of the CSF. The relation between H$_{0}$ and H$_{a0}$ can be further simplified and expressed as

\begin{equation}
    {\rm H}_{0} = {\rm H}_{a0}\left[1-\frac{d\Theta}{dz_{\rm FLRW}}(z_{\rm FLRW}=0) \right].
    \label{Hobs}
\end{equation}

Equation \ref{Dcz0} can be generalized to any redshift, and one can show that the observed Hubble parameter H$(z) = d(cz)/dD_{M}$ as measured, for example, from the BAO signal along the line of sight depends on the actual expansion rate H$_{a}(z_{\rm FLRW})$ in the following way:

\begin{equation}
    H(z_{\rm obs}) = H_{a}[z_{\rm FLRW}(z_{\rm obs})]\frac{dz_{\rm obs}}{dz_{\rm FLRW}}(z_{\rm obs}).
\end{equation}

Hereafter, H(z) and H$_{a}$(z) respectively refer to the observed Hubble parameter and the logarithmic expansion growth as defined above.

Having defined the observed Hubble parameter and the angular diameter distance, we could finally define the volume-averaged distance, which is a distance usually constrained by BAO measurements, 

\begin{equation}
    D_{\rm V}(z_{\rm obs}) = \left[cz_{\rm obs}(1+z_{\rm obs})^{2}D_{\rm A}^{2}(z_{\rm obs})H^{-1}(z_{\rm obs}) \right]^{1/3},
\end{equation}
and the Hubble distance,

\begin{equation}
    D_{\rm H}(z_{\rm obs}) = \frac{c}{H(z_{\rm obs})}.
\end{equation}

On the other hand, SNe Ia are also useful for calculating distances. Knowing the distance modulus, $\mu$, at a given redshift of a SN Ia, we can calculate the luminous distance, $D_{L}(z_{\rm obs})$, the following way:
\begin{equation}
    \mu = m(z_{\rm obs})-M(z_{\rm obs}) = 5\log_{10}[D_{L}(z_{\rm obs})]+25,
    \label{distancemodulii}
\end{equation}
where $m(z_{\rm obs})$ is the apparent magnitude, $M(z_{\rm obs})$ is the absolute magnitude, and $D_{L}(z_{\rm obs})$ is the luminosity distance at a given redshift:

\begin{equation}
    D_{L}(z_{\rm obs}) = (1+z_{\rm obs})D_{\rm M}(z_{\rm obs}) = (1+z_{\rm obs})^{2}D_{A}(z_{\rm obs}).
\end{equation}

\subsection{Cosmic microwave background}
The CMB temperature is linked to cosmic expansion through entropy, which remains conserved in our approximation since we do not modify the standard evolution of energy densities. Moreover, the radiation field has a vanishing trace, and its continuity equation retains its standard form, (see Eq. \ref{Tab_eq}). Inserting Eq. \ref{ocho} into this yields  $T \propto 1/(a\, e^{\Theta})$. Inserting Eq. \ref{eq8} into this relation yields  \( T_{\rm CMB} \propto (1 + z_{\rm obs}) \).

The redshift of decoupling is set by the ratio of the CMB temperature T$_{\rm dec}$ at decoupling to the CMB temperature T$_{0}$ measured at the present time, i.e.

\begin{equation}
    1+z_{\rm dec}=\frac{T_{\rm dec}}{T_{0}}.
\end{equation}

The current CMB temperature is known quite precisely from observations of the COBE-FIRAS \citep{1996ApJ...473..576F} satellite, yielding

\begin{equation}
    T_{0} = (2.7255\pm0.0006) \rm K.
\end{equation}
With T$_{\rm dec}\sim$3000K known from atomic physics, one gets z$_{\rm rec}\sim$1100. In the standard model with the FLRW metric, this redshift can be automatically converted into the cosmic scale factor at decoupling, $\displaystyle{a_{\rm dec}=\frac{1}{1+z_{\rm dec}}}$. This conversion, however, does not hold when we allow in the FLRW metric with CSF. In the latter case, the cosmic scale factor at decoupling also depends on the CSF:

\begin{equation}
    \frac{T_{\rm dec}}{T_{0}} = \frac{1}{a_{\rm dec}}\frac{1}{\exp{(\Theta_{\rm dec})}}.
\end{equation}

One way of incorporating the CSF into the analysis of the CMB observations is introducing a fiducial temperature of the CMB, 

\begin{equation}
    \hat{T}_{0} = T_{0}\exp{(-\Theta_{\rm dec})},
    \label{Tdec}
\end{equation}
 at the present time and holding the standard CMB computations with all atomic and particle data unchanged. This new fiducial temperature represents the present CMB temperature that would be measured by an imaginary instrument at $z_{\rm obs}=0$ in a rest frame with $\Theta=\Theta_{\rm dec}$ (from here on, we refer to this as the CSF frame). We therefore have to distinguish between all the (dimensional) parameters in the CSF frame (denoted by $\hat{x}$, where $x$ is the dimensional parameter) and the same parameters expressed in the local frame (where, by convention, $\Theta=0$). The transformation between these parameters in these two frames occurs in the following manner: In order to transform hat parameters to parameters expressed in the local frame, we had to scale all time and spatial dimensions expressed in the former frame by $\exp(-\Theta_{\rm dec})$. Therefore,  $\hat{H}_{a0}=H_{a0}\exp{(-\Theta_{\rm dec})}$ and $\hat{r}_{drag}=r_{\rm drag}\exp{(\Theta_{\rm dec})}$. 

\section{Parametrization of $\Theta(z_{\rm obs})$}\label{spline}
\subsection{Conformal function through the comoving distance equation}
One way of obtaining $\Theta(z_{\rm obs})$ is to require that these new cosmological models recover the actual distances in the Planck cosmology:

\begin{equation}
    \int_{0}^{z'}\frac{dz_{\rm FLRW}}{H_{a,new}(z_{\rm FLRW})} = \int_{0}^{z_{\rm obs}}\frac{dz_{\rm FLRW}}{H_{a,Planck}(z_{\rm FLRW})},
\end{equation}
where $z'$ is defined in equation \ref{zprime}. This equation (which must be solved numerically) allows one to relate $z_{\rm FLRW}$ and $z_{\rm obs}$ and, eventually, calculate $\Theta(z_{\rm obs})$ (see equation \ref{thetalog}).

\begin{figure}[htb!]
    \centering
    \includegraphics[width=0.45\textwidth]{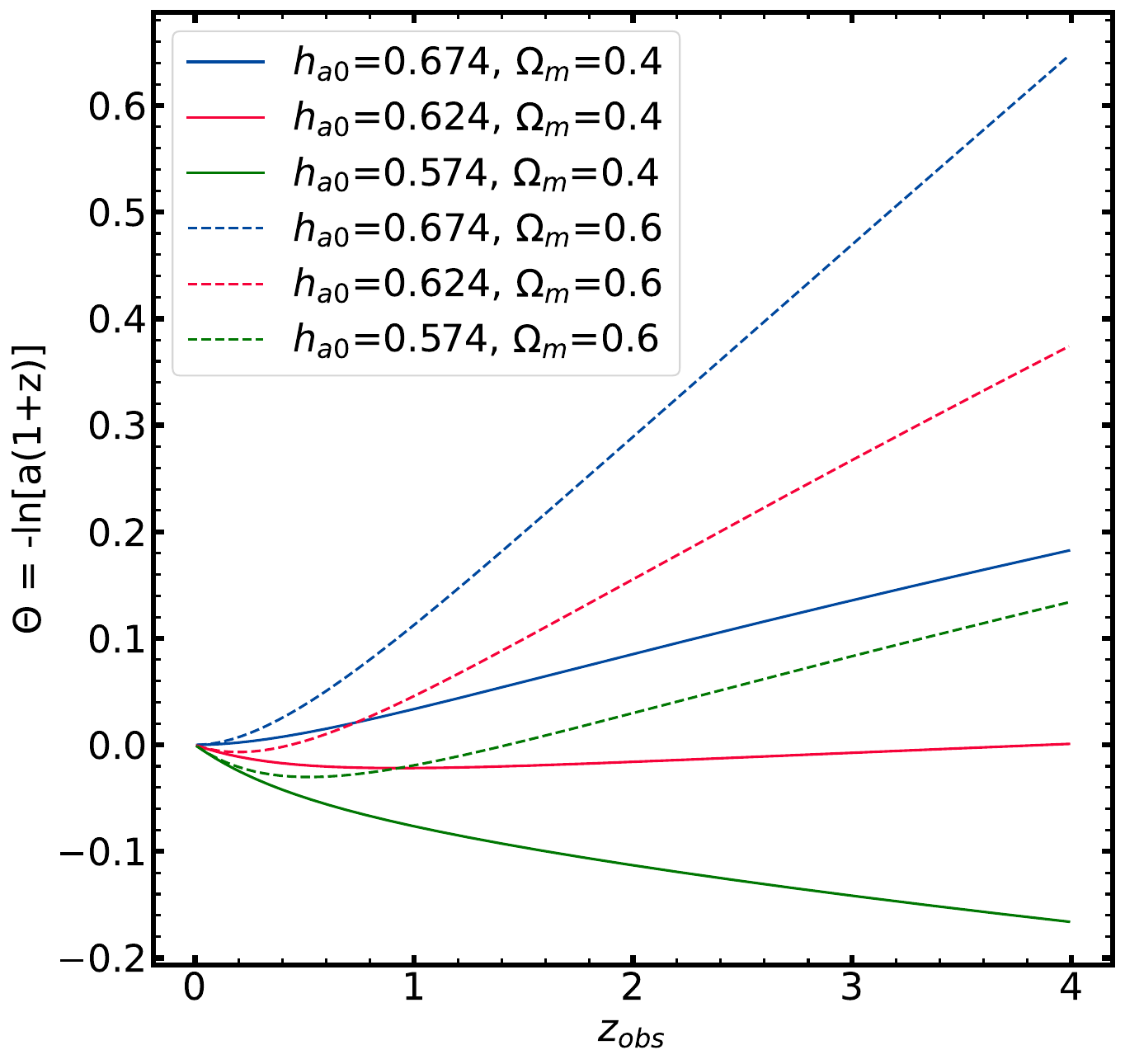}
    \caption{Redshift evolution of the CSF, $\Theta$, for a range of models with different matter density parameters, $\Omega_{\rm m}$, and the Hubble constant H$_{0}=h100$kms$^{-1}$Mpc$^{-1}$. All models recover distances and observed expansion history in the Planck cosmological model with $\Omega_{\rm m}$=0.315 and H$_{0}$=67.4 kms$^{-1}$Mpc$^{-1}$.}
    \label{theta_planck}
\end{figure}

In Figure \ref{theta_planck}, one can observe the redshift evolution of the CSF, $\Theta$, for a range of models with different density parameters, $\Omega_{\rm m}$, and the normalization of the Hubble constant H$_{0}=h100$kms$^{-1}$Mpc$^{-1}$. All of these models recover distances and observed expansion history in the Planck cosmological model with $\Omega_{\rm m}$=0.315 and H$_{0}$=67.4kms$^{-1}$Mpc$^{-1}$.

From Figure \ref{theta_planck}, we observed that when the new values of H$_{a0}$ or $\Omega{\rm m}$ differ significantly from the Planck values, $|\Theta| >> 1$ for high redshifts. However, if we want to maintain the same evolution for the density parameters as in the FLRW model, then $\Theta$ must be approximately constant, as mentioned in Section \ref{redshift_remapping}. We can avoid this behavior by restricting our considerations to models where the CSF, $\Theta$, stays within reasonable limits around $\Theta = 0$.

\begin{figure}[htb!]
    \centering
    \includegraphics[width=0.5\textwidth]{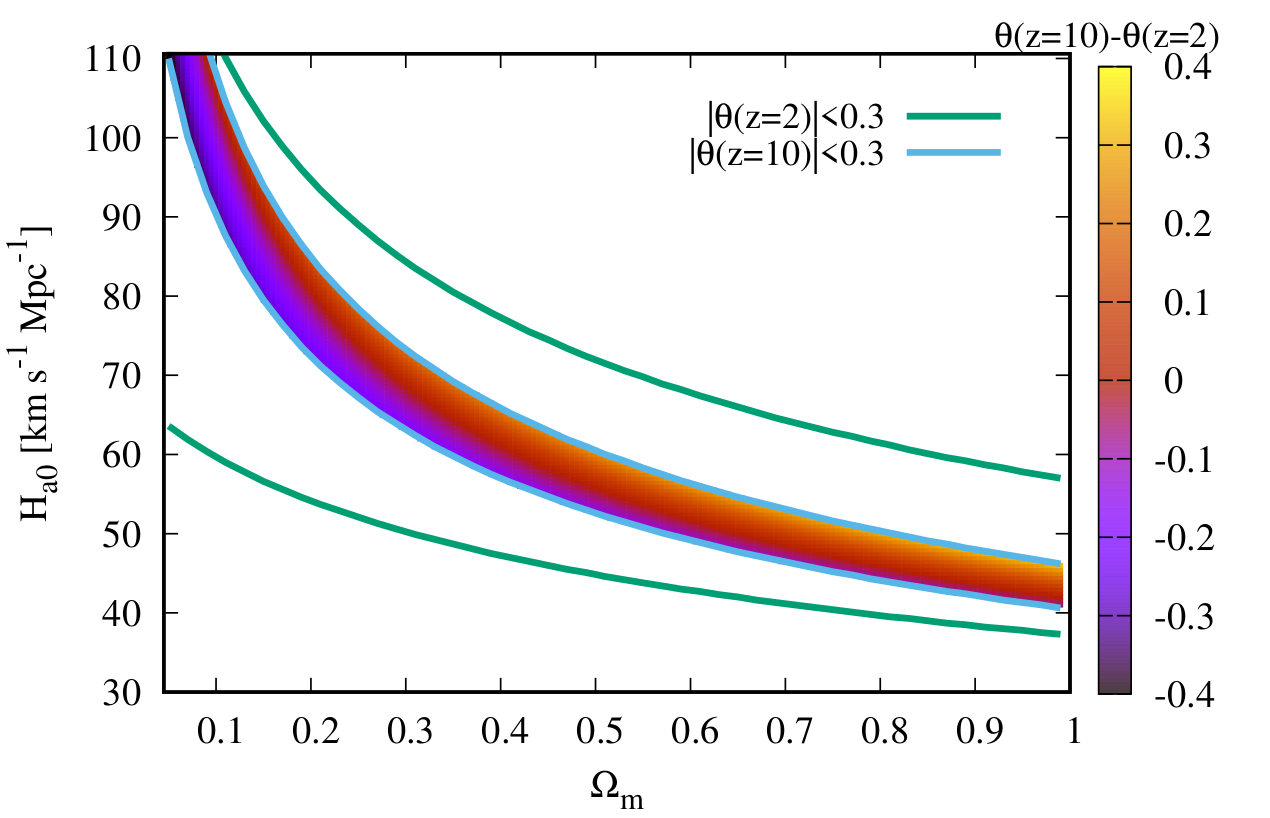}
    \caption{Bounds on the normalization of the Friedman equation H$_{a0}$ in CSF models found by imposing a $|\Theta|<0.3$ limit at $z<2$ and $z<10$. In this family of restricted solutions, cosmological models with deficient dark energy content require H$_{a0}<$H$_{0}$. The color map demonstrates the behavior of the $\Theta(z)$ profile at high redshifts as quantified by the difference in $\Theta$ between $z=10$ and $z=2$. The map shows that for every matter density parameter, there exists a unique solution for the CSF with $\Theta(z)=$ const at $z>2$.}
    \label{theta_planck_2}
\end{figure}

To investigate the allowed set of values for H$_{a0}$ and $\Omega{\rm m}$, we performed the following exercise: We checked for which values of H$_{a0}$ and $\Omega{\rm m}$ the conditions $|\Theta(z=2)|<0.3$ and $|\Theta(z=10)|<0.3$ are satisfied. This is shown in Figure \ref{theta_planck_2}, where these sets of values are delimited by a green line (for $|\Theta(z=2)|<0.3$) and a blue line (for $|\Theta(z=10)|<0.3$). Additionally, we use a color map to show the variation of $\Theta$ between these two redshift values for each pair of H$_{a0}$ and $\Omega{\rm m}$ values.

From Figure \ref{theta_planck_2}, one can see that the imposed limits on the CSF result in a clear trend for the preferred normalization of the Friedman equation, depending on the relative content of dark energy. In particular, models free of dark energy are characterized by H$_{a0} < $ H$_{0}$ and thus $\Theta < 0$ within a redshift range that includes at least the Hubble flow. Furthermore, we observed that for each value of $\Omega_{\rm m}$, there is a unique solution with a plateau (where $\Theta(z_{\rm obs})$ is constant for $z > 2$) in the profile of $\Theta(z_{\rm obs})$.

\subsection{Spline function}
An alternative way of obtaining $\Theta(z_{\rm obs})$ without making any theoretical presupposition is to assume a non-parametric form for this function for a redshift interval and to fit it with observational data, such as DESI-BAO Y1 results \citep{desicollaboration2024desi}. One possibility is to use a cubic spline function with four free nodes and derivatives in the extremes of this redshift interval. Therefore, we would have 4+2 free parameters regarding the spline function in addition to all the cosmological parameters involved in the distances and CMB that need to be found with observational data.

However, instead of using a cubic spline for $\Theta(z_{\rm obs})$, we used it for the ratio 

\begin{equation}
    \frac{z_{\rm FLRW}}{z_{\rm obs}} = 1+\alpha(z_{\rm obs}),
    \label{zflrw-zobs}
\end{equation}
as indicated in \cite{2017MNRAS.470.4493W}. The relationship between $\alpha(z_{\rm obs})$ and $\Theta(z_{\rm obs})$ is therefore

\begin{equation}
    \Theta(z_{\rm obs}) = \ln\left[\frac{1+z_{\rm obs} + \alpha(z_{\rm obs})z_{\rm obs}}{1+z_{\rm obs}} \right]
    \label{theta-alfa}.
\end{equation}

From equation \ref{theta-alfa}, one can see that $\Theta(z_{\rm obs}=0)$=0 independently of the value of $\alpha(z_{\rm obs}=0)$. 

Using this new parametrization (i.e., using $\alpha(z_{\rm obs})$ instead of $\Theta(z_{\rm obs})$) allowed us to compare our results with those obtained in \cite{2017MNRAS.470.4493W} more directly. Additionally, it has the advantage that, as shown in the aforementioned work, the range of $\alpha(z_{\rm obs})$ is much wider than the expected range of $\Theta(z_{\rm obs})$, so it will be easier to find the appropriate priors for the first case than for the second one.

\section{Datasets}\label{data}

\begin{table}[!tbh]
    \caption{Baryonic acoustic oscillation measurements from DESI Y1 \citep{desicollaboration2024desi} used in this work.}
    \centering
    \begin{tabular}{cccc}
        \toprule
         $z_{\rm eff}$&  Observable&  Measurement&  Tracers\\
         \midrule
 0.30& $D_{\rm V}(z)/r_{\rm drag}$& 7.93 $\pm$ 0.15&DESI BGS\\
 0.51& $D_{\rm M}(z)/r_{\rm drag}$& 13.62 $\pm$ 0.25&DESI LRG\\
 0.51& $D_{\rm H}(z)/r_{\rm drag}$& 20.98 $\pm$ 0.61&DESI LRG\\
  0.71& $D_{\rm M}(z)/r_{\rm drag}$& 16.85 $\pm$ 0.32& DESI LRG\\
 0.71& $D_{\rm H}(z)/r_{\rm drag}$& 20.08 $\pm$ 0.60& DESI LRG\\
 0.93& $D_{\rm M}(z)/r_{\rm drag}$& 21.71 $\pm$ 0.28& DESI LRG+ELG\\
 0.93& $D_{\rm H}(z)/r_{\rm drag}$& 17.88 $\pm$ 0.35& DESI LRG+ELG\\
 1.32& $D_{\rm M}(z)/r_{\rm drag}$& 27.79 $\pm$ 0.69& DESI ELG\\
 1.32& $D_{\rm H}(z)/r_{\rm drag}$& 13.82 $\pm$ 0.42& DESI ELG\\
 1.49& $D_{\rm V}(z)/r_{\rm drag}$& 26.07 $\pm$ 0.67& DESI QSO\\
 2.33 & $D_{\rm M}(z)/r_{\rm drag}$& 39.71 $\pm$ 0.94& DESI Ly $\alpha$ QSO\\
 2.33 & $D_{\rm H}(z)/r_{\rm drag}$& 8.52 $\pm$ 0.17& DESI Ly $\alpha$ QSO\\
 \bottomrule
    \end{tabular}
    \label{BAO_data}
\end{table}

As mentioned in the previous section, this work uses observational data from BAO distances and the Hubble parameter, distance moduli from SNe Ia, and the CMB temperature power spectrum.
In Table \ref{BAO_data}, we present the BAO measurements from DESI \citep{desicollaboration2024desi} for galaxy, quasar, and Lyman-$\alpha$ forest tracers from the first year of observations compiled in the DESI Data Release 1. These BAO measurements consist of $D_{\rm M}$, $D_{\rm H}$, and $D_{\rm V}$ at the effective redshifts of $z_{\rm eff} = 0.30, 0.51, 0.71, 0.92, 1.23, 1.49, 2.33$.

The SN Ia dataset used in this work is the Pantheon+ compilation \citep{2022ApJ...938..110B}, which consists of 1550 spectroscopically confirmed SNe Ia in the redshift range $0.001 < z_{\rm obs} < 2.26$. However, we only considered supernovae with redshifts $z_{\rm obs} \geq 0.01$ in order to mitigate the impact of peculiar velocities in the Hubble diagram \citep{2022ApJ...938..112P}. It is important to note that SNe Ia alone cannot constrain H$_{0}$ unless they are combined with other datasets \citep{2022ApJ...938..110B, 2024ApJ...973L..14A}. Therefore, we did not attempt to constrain the new parameters using Pantheon+ alone and instead used DESI, CMB, and Pantheon+ data together; i.e., instead of calibrating distance using the distance ladder, we calibrated it using the sound horizon.

Additionally, we used the sample of SNe Ia from the Dark Energy Survey (DES), part of their Year 5 data release (DESY5). This dataset contains 1635 photometrically classified SNe Ia with redshifts $0.1 < z_{\rm obs} < 1.4$ complemented by 194 low-redshift SNe Ia (which overlap with the Pantheon+ sample) spanning $0.025 < z_{\rm obs} < 0.1$ \citep{2024ApJ...973L..14A}. We used the two datasets interchangeably in our analysis, but they were not combined in the same global likelihood due to partial correlations.

All the data presented so far consist of low-redshift probes (LZ). However, to ensure that $\Theta(z_{\rm obs})$ remains nearly constant for high values of $z_{\rm obs}$, we included the CMB and obtained the value of $\Theta(z_{\rm dec})$ from equation \ref{Tdec}. This was achieved by treating the CMB temperature, $T_{\rm dec}$, as a free parameter.

The CMB temperature power spectrum can be computed using the \texttt{CAMB} code \citep{2000ApJ...538..473L}, which depends on the following cosmological parameters: H$_{a0}$, $\Omega{\rm b} h_{a}^2$, $\Omega_{\rm c} h_{a}^2$, $M_{\nu}$, $\Omega_{\rm k}$, $\tau$, $T_{\rm obs0}$, $A_{\rm lens}$, $A_{\rm s}$, $n_{\rm s}$, and $r$, which respectively represent the Hubble constant, the physical baryon density parameter, the physical dark matter density parameter, the sum of the three neutrino masses, the curvature density parameter, the reionization optical depth, the CMB temperature, the dimensionless lensing parameter, the primordial amplitude of scalar perturbations (spectral index), and the ratio of tensor to scalar perturbation amplitudes.
However, caution must be exercised when using \texttt{CAMB} to compute the CMB power spectrum in the context of the CSF model. As mentioned earlier, in this case the CMB temperature must be treated as a free parameter, which can be related to the observed CMB temperature by COBE through equation \ref{Tdec}. Additionally, since conformal transformations, and the CSF model in particular, preserve angles, the power spectra in dimensionless normalization (i.e., for $\Delta T/T$ rather than $\Delta T$) obtained by an observer at the decoupling epoch in the CSF framework and modulo standard expansion effects as well as secondary CMB effects should remain consistent.

However, we note that if the CMB temperature is treated as a free parameter and \texttt{CAMB} is used, all cosmological parameters will be expressed in the frame of reference of an observer at the decoupling epoch, whereas the cosmological parameters appearing in the BAO distance equations are expressed in the present epoch. To transform the former to the latter, we had to scale all time and spatial dimensions by $\exp(-\Theta_{\rm dec})$--all dimensionless quantities remain the same regardless of the frame of reference.

\section{Methods}\label{methods}
We used Markov chain Monte Carlo technique to find the best-fitting parameters of each model. The likelihood function for the BAO and SN Ia datasets is given by
\begin{equation}
    \ln{\mathcal{L}(\bm{\Theta}| {\rm data})} \propto -\frac{\chi^{2}_{\rm SN}}{2}-\frac{\chi^{2}_{\rm BAO}}{2},
\end{equation}
where $\chi^{2}$ is calculated using the chi square function, which in both cases is calculated as
\begin{equation}
    \chi^{2} = \sum_{\rm i,j}^{N}C_{\rm ij}^{-1}[m_{\rm i}-m(z_{\rm obs,i},\bm{\Theta})][m_{\rm j}-m(z_{\rm obs, 
    j},\bm{\Theta})],
\end{equation}
where $\bm{\Theta}$ is a vector of the model parameters and $C$ is the covariance matrix. 
The likelihood function used for the CMB temperature power spectrum includes all power spectra and lensing corresponding to the Planck 2018 data release \citep{planck2018}.

We used the Markov chain Monte Carlo sampler from \texttt{Cobaya} \citep{2019ascl.soft10019T, Torrado_2021} and set uniform priors for all free parameters related to BAO distances and magnitude distances. These priors are listed in Table \ref{priors}. Chain convergence was assessed using the Gelman-Rubin test \citep{10.1214/ss/1177011136}. Once a tolerance of 0.01 was achieved, the chains were considered to have converged. Additionally, we discarded the first 30$\%$ of each chain as burn-in and used the mean value along with the 68$\%$ two-tail equal-area confidence limit to represent the best-fit value and its uncertainty.

\begin{table}[htb!]
    \caption{Prior choices made in this work.}
    \centering
    \begin{tabular}{cc}
    \toprule
         Parameter& Priors\\
         \midrule
         $h_a=H_{0a}/(100 km/s/Mpc)$&   [0.1-1.2]\\
         $\alpha(0.0)$&  [-0.7-0.01] \\
         $\alpha(0.5)$&  [-0.7-0.01] \\
         $\alpha(1.0)$&  [-0.7-0.01] \\
         $\alpha(2.5)$&  [-0.7-0.01] \\
         $d\alpha/dz_{\rm obs}(0.0)$  & [0.05-0.3] \\
         $d\alpha/dz_{\rm obs}(2.5)$  & [0-0.1]\\
         $\Omega_{\rm m}$ &   [0.1-0.99]\\
         $\Omega_{\Lambda}$&  1-$\Omega_{\rm m}$\\
         $r_{\rm drag}/$Mpc&  [50-200]\\
         \bottomrule
    \end{tabular}
    \label{priors}
\end{table}

For the calculation of the CMB, we sampled the following parameters: $\hat{T}_{0}$, $\log(A)$, $n_{s}$, $\hat{\rm H}_{a0}$, $\Omega_{b}\hat{h}_{a}^{2}$, $\Omega_{c}\hat{h}_{a}^{2}$, $\tau$, and $\hat{r}_{\rm drag}$. Gaussian priors were set for each parameter, centred on the Planck best-fit values, except for $\hat{T}_{0}$, for which we used a uniform prior in the range [2.5, 4].

\section{Results}\label{results}
\begin{figure}[htb!]
    \centering
    \includegraphics[width=0.5\textwidth]{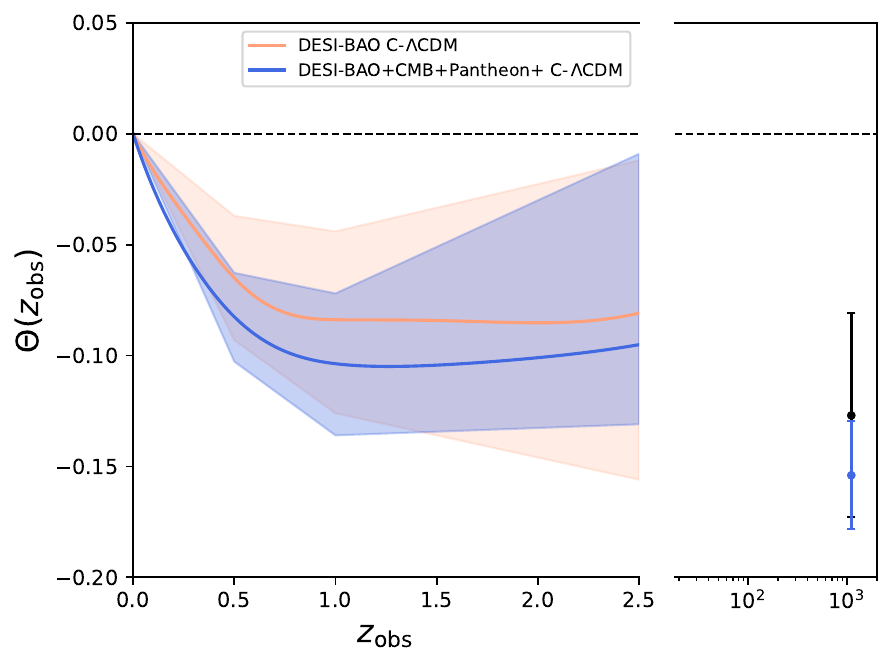}
    \caption{For a flat CSF-$\Lambda$CDM model, $\Theta(z_{\rm obs})$ inferred from DESI-BAO Y1, Pantheon+, and DESI+CMB+Pantheon+. The shadow region delimits the 68$\%$ uncertainty. The black and blue points represent the values of $\Theta$ at recombination, $\Theta_{\rm dec}$, obtained fitting CMB and DESI+CMB+Pantheon+, respectively, in the CSF model. }
    \label{theta_desi}
\end{figure}

We begin this section by considering a flat Universe with dark energy characterized by the condition $\Omega_{\rm m} + \Omega_{\Lambda} = 1$. In Figure \ref{theta_desi}, we present the function $\Theta(z_{\rm obs})$ inferred from DESI-BAO distance measurements for a flat C-$\Lambda$CDM model alongside the results from the combined DESI-BAO Y1 distances, CMB, and Pantheon+ (DESI+CMB+Pantheon+). The shaded region represents the 68$\%$ confidence intervals. We observed that the CSF inferred from DESI-BAO is more positive compared to the one obtained from DESI+CMB+Pantheon+. Additionally, the best-fit value of $\Theta(z_{\rm dec})$ obtained from fitting DESI+CMB+Pantheon+ is more negative than the one derived from the CMB alone, although the value is still compatible with that from the CMB within 1$\sigma$.

An important point to note is that the CSF model, when fitted to the DESI-only or DESI+CMB+Pantheon+ datasets, is not compatible with $\Theta(z_{\rm obs}) = 0$ within 1$\sigma$. For the best-fit values presented in the second column of Table \ref{open_values}, we find that the value of $-\log{\mathcal{P}}$ is 3.25, where $\mathcal{P}$ denotes the posterior distribution. For the $\Lambda$CDM model, we obtained $-\log{\mathcal{P}} = 11.53$. When fitting only the BAO data, we have two free parameters ($\Omega_{\rm m}$ and $h_{a}r_{\rm d}$) and 12 data points (see Table \ref{BAO_data}). Therefore, for the standard model, we have 10 degrees of freedom, and for the CSF model, we have 4 degrees of freedom (12-2-6, where 6 comes from the six free parameters of the spline function). Consequently, the reduced chi-square values are $\chi^{2}_{\nu}$ = 2.3 for the standard model and $\chi^{2}_{\nu}$ = 1.6 for the CSF model. This indicates that the CSF model provides a better fit to the data, as it has a lower reduced chi-square value.

We can also compare the $\chi^{2}$ values when fitting DESI+CMB+Pantheon+, although calculating $\chi^{2}_{\nu}$ is more complicated, as calculating the number of degrees of freedom from the CMB is a hard task. Nevertheless, $-\log{\mathcal{P}}=2102.66$ for the CSF model (where $\mathcal{P}$ is the posterior that corresponds to the best-fit values from the Table \ref{open_values}), and $-\log{\mathcal{P}}=2121.1993$. These two values lead to a $\chi^{2} = -2\Delta\log{\mathcal{L}} = 37.1$, which also implies an improvement in the CSF model with respect to the standard model. However, in this $\chi^{2}$ value we do not take into account the difference in the number of degrees of freedom. To do so, we can apply the Akaike information criterion (AIC; \citep{1974ITAC...19..716A}):

\begin{equation}
    \rm{AIC} = 2k + 2(-\log{\mathcal{L}_{\rm max}}),
\end{equation}

where $k$ is the number of free parameters and $\mathcal{L}_{\rm max}$ is the maximum of the likelihood.

For the standard $\Lambda$CDM model, we obtained AIC$_{\rm standard}$ = 4254.3986, while for the CSF model, AIC$_{\rm CSF}$ = 4231.31282. Given these AIC values, the CSF model is favored, as it has a lower AIC value, indicating a better balance between goodness of fit and model complexity (despite having more free parameters).

Focusing again on Figure \ref{theta_desi}, we also observed that for $z_{\rm obs} > 0.5$, $\Theta(z_{\rm obs})$ remains approximately constant when fitting only DESI and DESI+CMB+Pantheon+. Furthermore, when considering the value provided by the CMB at the decoupling epoch (represented by the black point in Figure \ref{theta_desi}), we found that the CSF remains nearly constant for $z_{\rm obs} > 2.5$ until $\Theta_{\rm dec}$, as discussed in Section \ref{redshift_remapping}.

The next step was to examine whether approximating the evolution of the density parameters using the FLRW metric is a reasonable approach for the CSF model. In other words, we needed to assess whether $\partial \Theta / \partial a$ in equation \ref{rho_eq} is negligible. From the DESI+CMB+Pantheon+ column in Table \ref{open_values}, we observed that $\Theta(2.5) = -0.096$ for DESI+CMB+Pantheon+, which results in a value of $\exp(\Theta_{0} - \Theta(2.5)) = \exp(0.096) \sim 1.1$. This suggests that we overestimated the evolution of the density parameters by approximately 10$\%$. However, the uncertainties in this work are large enough to justify this approximation.

\begin{figure*}[htb!]
    \centering
    \includegraphics[width=0.495\textwidth]{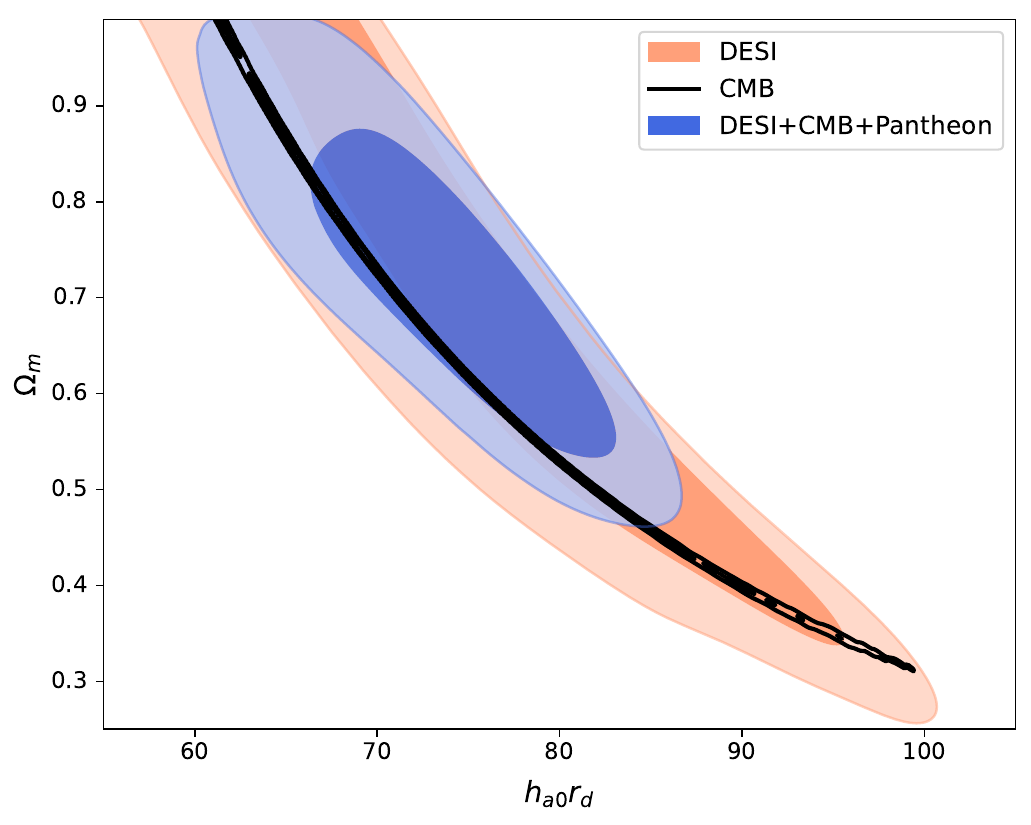}
    \includegraphics[width=0.49\textwidth]{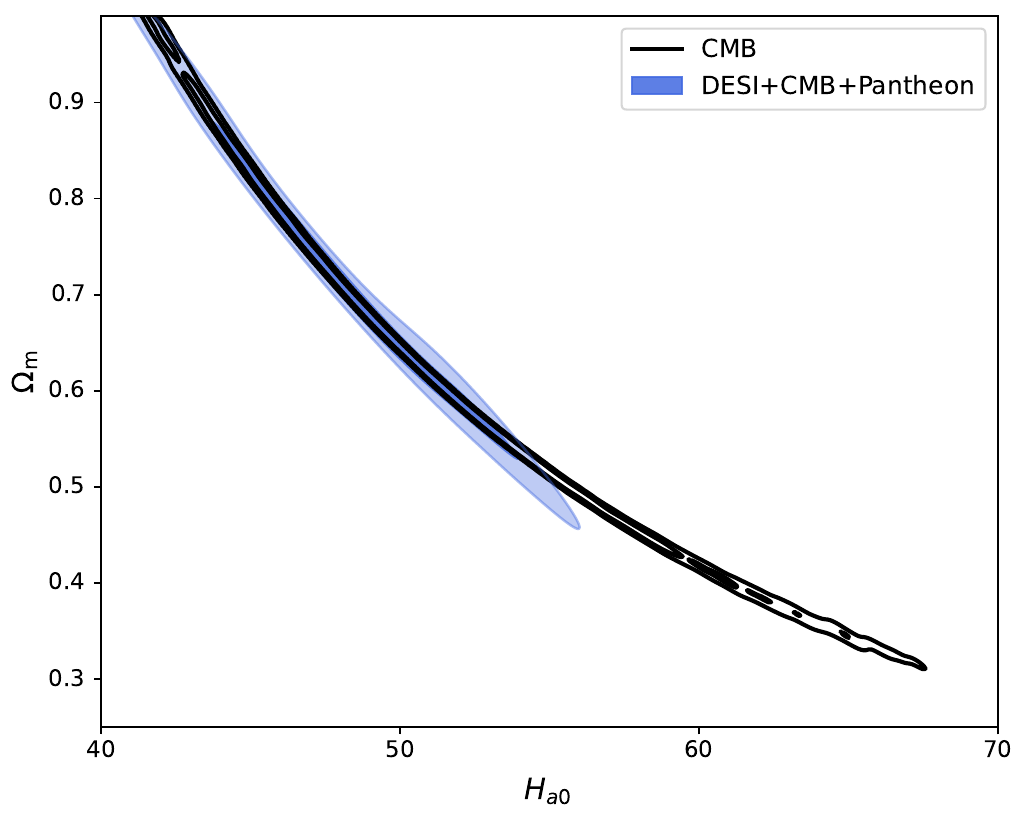}
    \caption{Left: Marginalized 68\% and 95\% confidence level constraints on the matter density parameter, $\Omega_{\rm m}$, and the Hubble constant, H$_{0}$, multiplied by the sound horizon scale, $r_{\rm drag}$, assuming a CSF flat $\Lambda$CDM model using only DESI-BAO Y1, only Pantheon+, and DESI+CMB, Pantheon+. Right: Same as in the left panel but for the $\Omega_{\rm m}-$H$_{a0}$ plane.}
    \label{desi_contour}
\end{figure*}

The confidence level contours in the $\Omega_{\rm m}$-$h_{a}r_{\rm drag}$ (and $\Omega_{\rm m}$-H$_{a0}$) plane are shown in the left (right) panel of Figure \ref{desi_contour}. The best-fit values of all the parameters are provided in Table \ref{open_values}. As shown in Figure \ref{theta_desi}, the constraints on the CSF from DESI and DESI+CMB+Pantheon+ are compatible within 1$\sigma$.

Regarding $\Omega_{\rm m}$, we found a higher value than the one provided by Planck for the flat $\Lambda$CDM model. We also observed that DESI and CMB yield similar results for $\Omega_{\rm m}$, but the combination of DESI+CMB+Pantheon+ gives a higher value, likely influenced by Pantheon+.

We also compared the value of H$_{0}$ (the observed Hubble constant) obtained from the CSF model with that from the standard model. DESI-BAO cannot constrain H$_{a0}$ or H$_{0}$, and the CMB can only constrain H$_{a0}$ (the derivative of $\Theta(z_{\rm obs})$ at $z_{\rm obs} = 0$ is needed to obtain H$_{0}$ from H$_{a0}$, as illustrated in equation \ref{Hobs}). However, the combination of DESI+CMB+Pantheon+ can constrain both H$_{a0}$ and H$_{0}$ simultaneously because DESI constrains the CSF, and the CMB provides the Hubble constant from the standard metric, H$_{a0}$. The value of the observed Hubble constant constrained from DESI+CMB+Pantheon+ is H$_{0}$ = 64.03$\pm$3.39, which is compatible with the Planck 2018 result \citep{planck2018} (H$_{0}$ = 67.36$\pm$0.54) within 1$\sigma$.

Next, we focus on $\sigma_{8}$. We observed that neither of the two values (the one obtained from the CMB and from DESI+CMB+Pantheon+ in the CSF model) is compatible with the value given by the standard model fitting DESI+CMB+Pantheon+ or with Planck's value ($\sigma_{8} = 0.8111 \pm 0.0060$). However, this is not a problem because the power spectrum is preserved. It is important to note, however, that caution is required when comparing the matter power spectrum with the one provided by the standard model. If we wish to represent the matter power spectrum obtained in the CSF model in terms of $\Theta = 0$, as observed in the standard metric, this power spectrum must be transformed by multiplying it by $\exp{(-3\Theta_{\rm dec})}$ since the units of the power spectrum are in distance units raised to the power of three. Additionally, we must multiply by the ratio $h_{\rm Planck}/h_{\rm a0}$, where $h_{\rm Planck}$ is the reduced Hubble constant value from Planck in the standard metric, and $h_{\rm a0}$ is the Hubble constant in the CSF model. This adjustment is necessary due to the differing units of the matter power spectrum.

The next important parameter to study was the age of the Universe, which is presented in the last row of Table \ref{open_values}. We observed that the value obtained for a standard flat $\Lambda$CDM model is 13.78 $\pm$ 0.02 billion years. However, for a CSF flat $\Lambda$CDM model, the predicted age of the Universe is smaller: 13.11 $\pm$ 0.13 billion years. The tension between these two values is approximately 5$\sigma$. This decrease in the age of the Universe is a direct consequence of the increase in the matter content and decrease in dark energy within the Universe.

We can compare the obtained age of the Universe with that from very old objects, such as some stars or globular clusters. For example, one finds the ages of the oldest stars 2MASS J18082002–5104378 B equal
to t* = 13.535$\pm$0.002 Gyr \citep{2018ApJ...867...98S}, but if the scatter among different models to fit for the age is taken into account,
the age becomes t* = 13.0 $\pm$ 0.6 Gyr \citep{2019JCAP...03..043J}. Furthermore, the age of HD 140283 that is equal to t* = 14.46 $\pm$ 0.8 Gyr \citep{2013ApJ...765L..12B}
becomes t*= 13.5$\pm$0.7 Gyr when using the new Gaia parallaxes instead of the original HST parallaxes \citep{2021APh...13102607D}. Therefore, taking into account the scatter among different models to fit for the age, the age of the Universe we obtain in this work is compatible within 1$\sigma$ with the ages of these stars.

\begin{table*}[htb!]
\caption{Posterior mean and standard deviation for the parameters of the CSF flat $\Lambda$CDM model from DESI-BAO Y1, CMB, and Pantheon+ data compared with the standard $\Lambda$CDM model.}
    \centering
    \renewcommand{\arraystretch}{1.25} 
    \begin{tabular}{c|cccc}
    \toprule
         Parameter&  DESI  &CMB&DESI+CMB+Pantheon+&Base-$\Lambda$CDM\\
         \midrule
         $100\Omega_{\rm b}h^{2}$&  -- &2.555$\pm$0.122&2.622$\pm$0.081&2.248$\pm$0.013\\
         $\Omega_{\rm c}h^{2}$&  -- &0.135$\pm$0.006&0.137$\pm$0.003&0.1186$\pm$0.0008\\
         $\tau$&  -- &0.0515$\pm$0.0076&0.0512$\pm$0.0073&0.058$\pm$0.007\\
         ln($10^{10}A_{\rm s}$)& -- &3.040$\pm$0.015&3.038$\pm$0.015&3.049$\pm$0.015\\
         $n_{s}$&  -- &0.9675$\pm$0.0043&0.9696$\pm$0.0056&0.9685$\pm$0.0035\\

         $\Theta_{\rm dec}$&  -- &-0.127$\pm$0.046&-0.148$\pm$0.026&0\\

         $\Theta(0.5)$& -0.065$\pm$0.028&--&-0.081$\pm$0.020&0\\
         $\Theta(1.0)$& -0.085$\pm$0.041&--&-0.102$\pm$0.032&0\\
         $\Theta(2.5)$& -0.084$\pm$0.072&--&-0.094$\pm$0.061&0\\
         $d\Theta/dz_{\rm obs}$(0.0)& -0.192$\pm$0.092&--&-0.275$\pm$0.056&0\\
         $d\Theta/dz_{\rm obs}$(2.5)& 0.0202$\pm$0.024&--&0.016$\pm$0.021&0\\
         H$_{a0}$& -- &50.25$\pm$6.04&47.77$\pm$3.12&67.99$\pm$0.39\\
        \midrule
         H$_{0}$& -- &--&64.05$\pm$3.45&H$_{a0}$\\
         $\Omega_{\rm m}$& 0.655$\pm$0.189&0.672$\pm$0.173&0.729$\pm$0.111&0.3067$\pm$0.0051\\
         $h_{a}r_{d}$& 77.41$\pm$9.20&74.01$\pm$8.85&73.10$\pm$5.39&100.18$\pm$0.66\\

         $\sigma_{8}$&  -- &0.667$\pm$0.055&0.642$\pm$0.032&0.809$\pm$0.006\\
        $r_{d}/$Mpc& -- &147.27$\pm$0.28&153.04$\pm$6.02&147.34$\pm$0.21\\
        Age [Gyr] & -- & 13.114$\pm$0.266& 13.11$\pm$0.13&13.78$\pm$0.02\\
        \bottomrule
        \end{tabular}
    \tablefoot{Posterior mean and standard deviation for the parameters of CSF flat $\Lambda$CDM model. The first column shows the constraints from using only DESI-BAO Y1 measurements, the second column shows the constraints for only CMB, the third column shows the constraints obtained combining DESI-BAO Y1 measurements with CMB and Pantheon+, and the last column shows the constraints from DESI+CMB+Pantheon+ from the standard base $\Lambda$CDM model. The parameters above the middle line represent the fundamental parameters of each model (6+6 parameters for the CSF models, and six parameters for the standard model).}
    \label{open_values}
\end{table*}

The next step was to examine how the CSF model affects various power spectra. For this analysis, we considered the constrained parameters obtained by fitting only the CMB with the CSF metric.

\begin{figure}[htb!]
    \centering
    \includegraphics[width=0.5\textwidth]{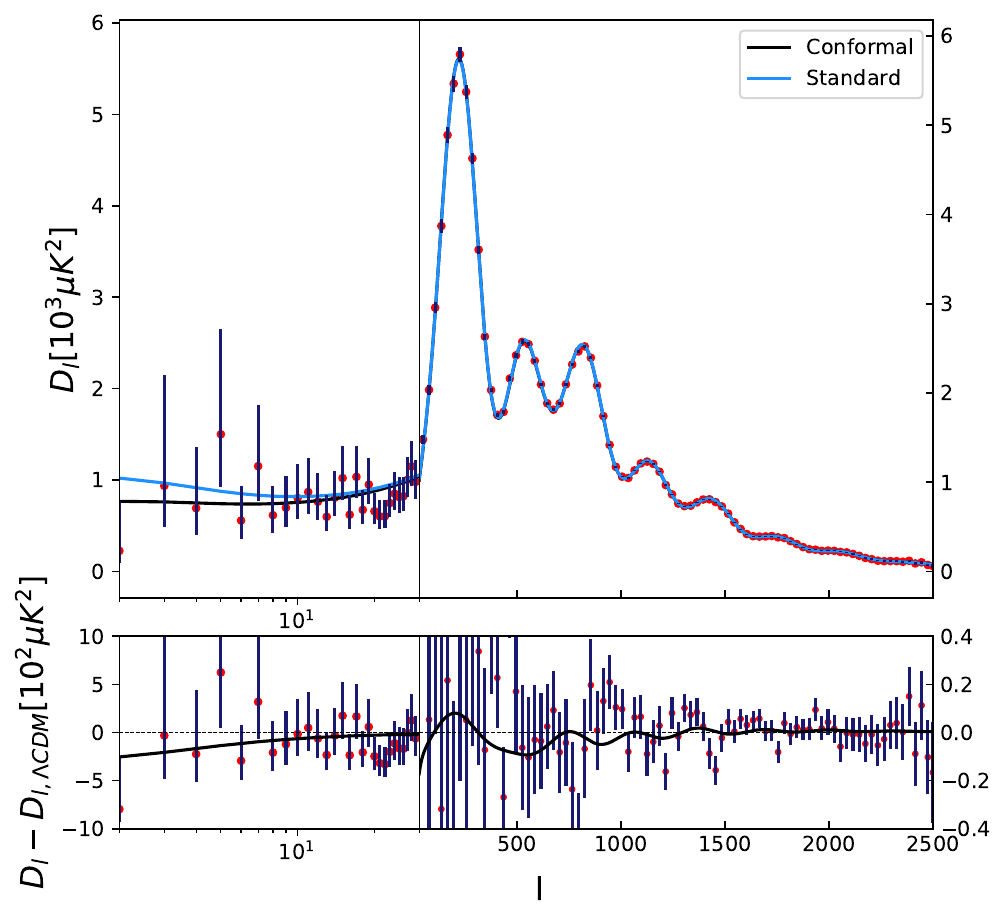}
    \caption{Top panel: Temperature power spectrum obtained with the standard model using Planck 2018 best-fit values (blue line) and with the CSF model using the best-fit values obtained fitting only the CMB (red line). The red points are the values measured by Planck 2018 \citep{planck2018}. Bottom panel: Difference between the CSF and standard model. } 
    \label{TT_CMB}
\end{figure}

In Figure \ref{TT_CMB}, we show the temperature power spectrum obtained using the FLRW (standard) metric with Planck 2018 cosmological parameters and the one obtained with the CSF metric using the constrained parameters.\footnote{We note that if we want the temperature power spectrum in the CSF metric to be represented at $\Theta=0$, as in the observations, this power spectrum must be transformed by multiplying it by $\exp{(2\Theta_{\rm dec})}$ since the units of the power spectrum are temperature squared.} The main difference between the two spectra is observed at low values of $l$, which could have an important interpretation, namely, a decrease in the integrated Sachs-Wolfe effect \citep{1967ApJ...147...73S, 1968Natur.217..511R, 1990ApJ...355L...5M, 1995ApJS..100..281S}.

\begin{figure}[htb!]
    \centering
    \includegraphics[width=0.5\textwidth]{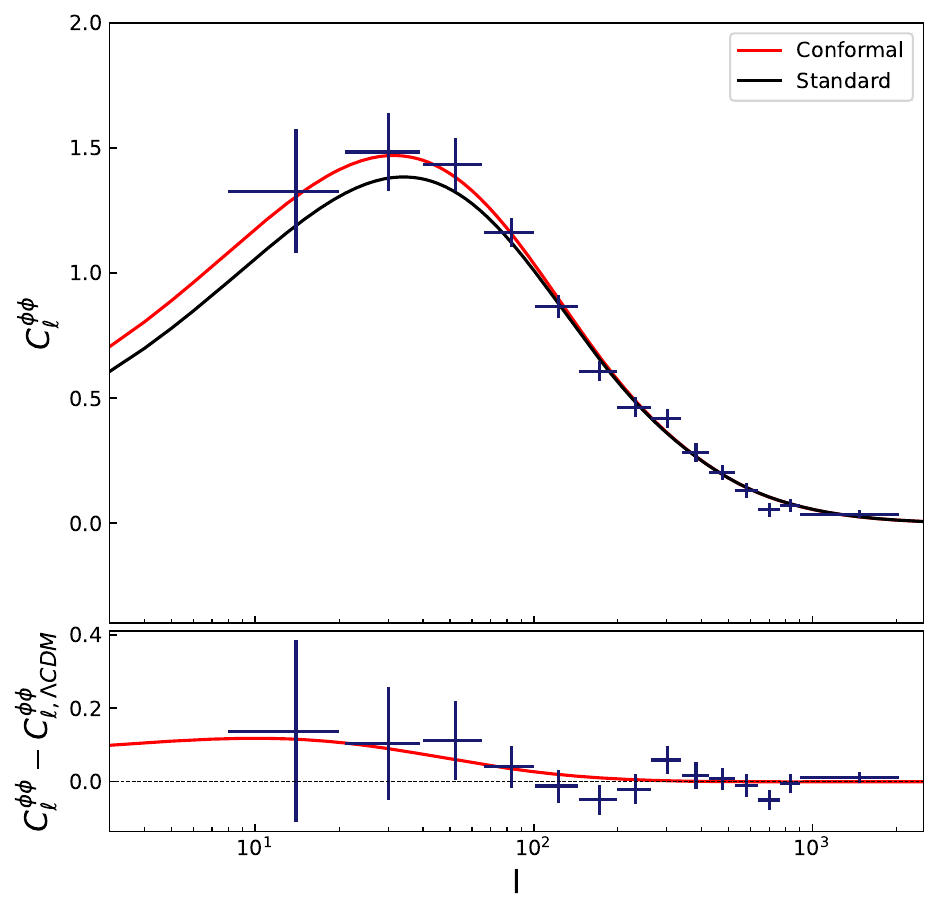}
    \caption{Top panel: Lensing potential power spectrum obtained with the standard model using Planck 2018 best-fit values (black line) and with the CSF model using the best-fit values obtained fitting only the CMB (red line). The red points are the values measured by Planck 2018 \citep{planck2018}. Bottom panel: Difference between the CSF and standard model. } 
    \label{lensing_CMB}
\end{figure}

In Figure \ref{lensing_CMB}, we show the lensing potential power spectrum obtained from both the CSF (red line) and standard (black line) models. The observational data measured by Planck \citep{planck2018} is also included for comparison. It is evident that the CSF model provides a better fit to the observational data for $C_{l}^{\phi\phi}$ at low values of $l$, although the uncertainties are large, and the observational data is compatible with both models.

\begin{figure*}
    \centering
    \includegraphics[scale=0.5]{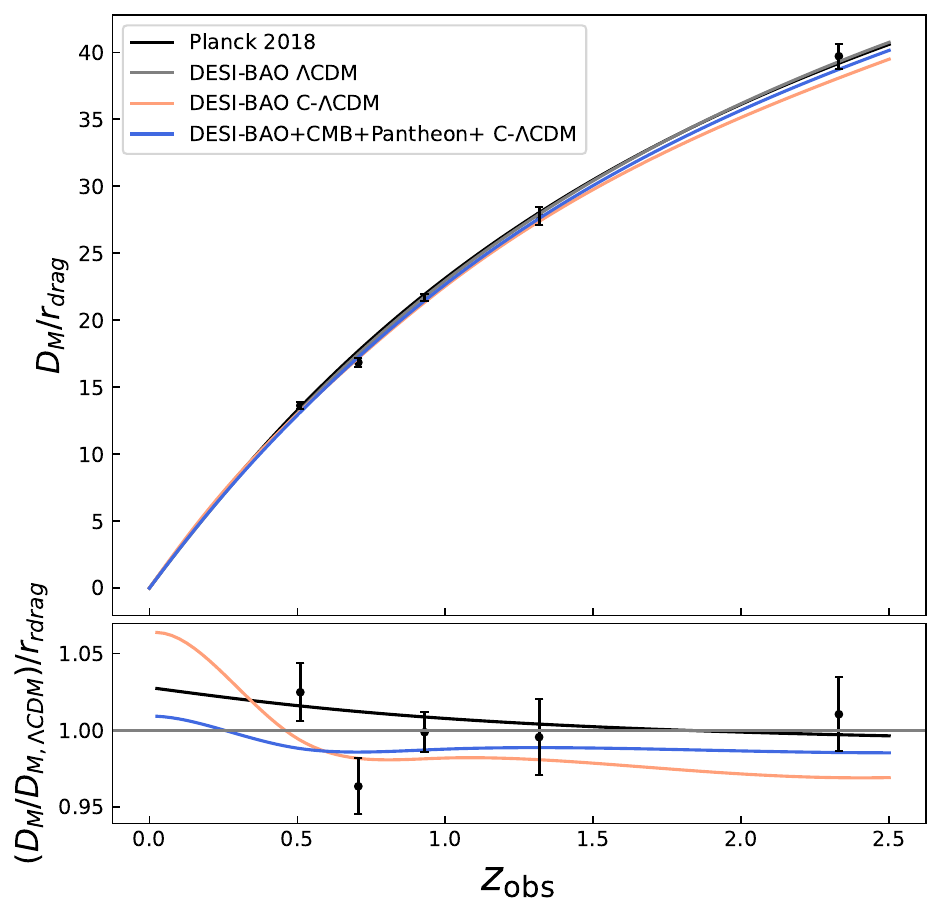}
    \includegraphics[scale=0.5]{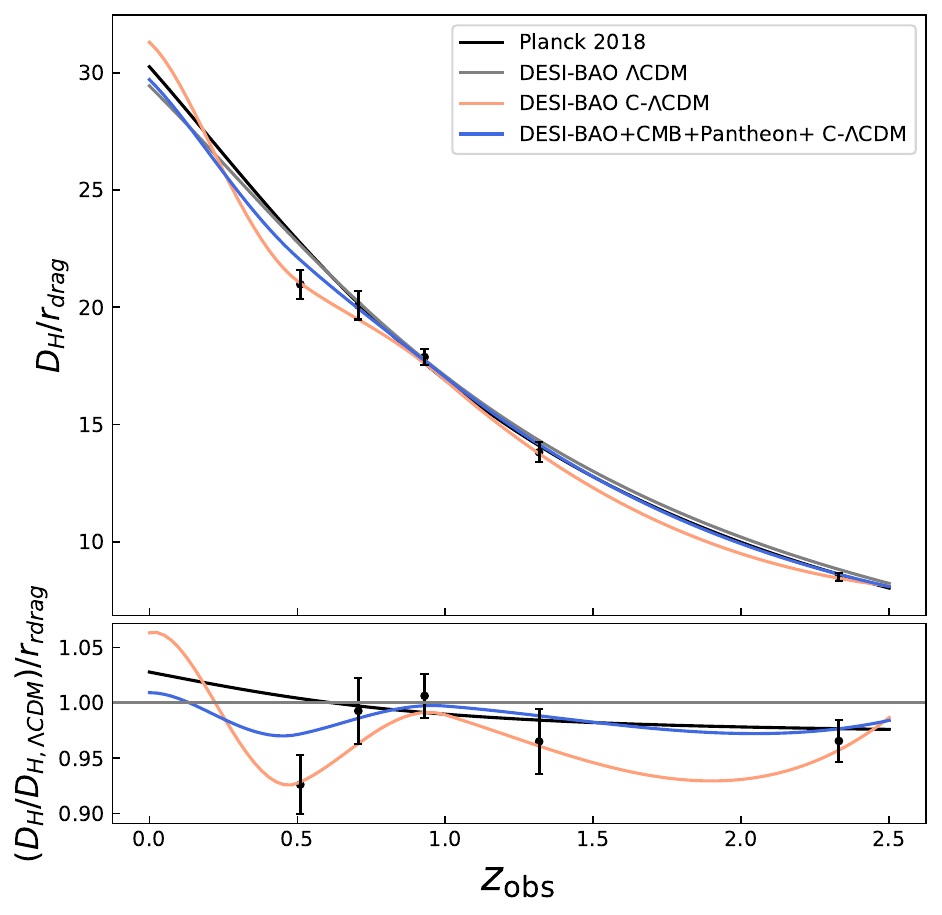}\\
    \includegraphics[scale=0.5]{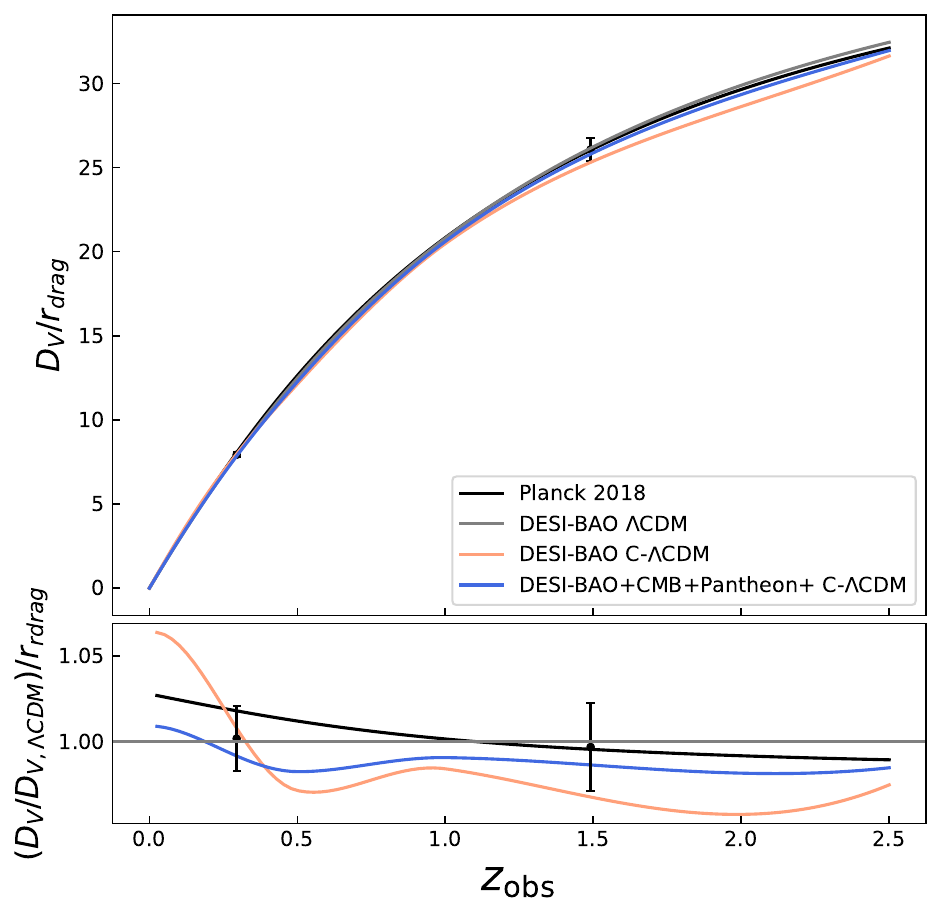}
    \includegraphics[scale=0.5]{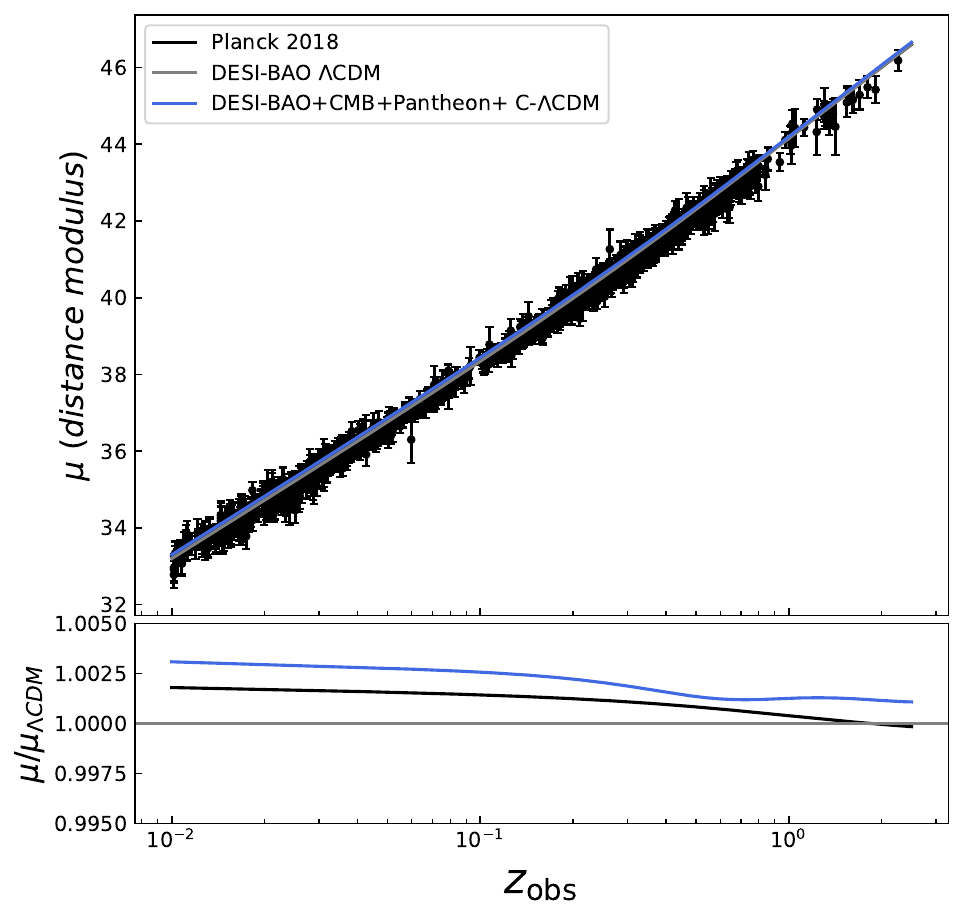}\\
    \caption{Top left: Obtained fitting observations (black points) of  $D_{M}/r_{\rm drag}$ to CSF models using DESI-BAO and DESI-BAO+CMB+Pantheon+. We also show the result of fitting only DESI-BAO to a standard $\Lambda$CDM model and the results obtained with Planck 2018 parameters \citep{planck2018} for a standard $\Lambda$CDM model. Top right: Same but for $D_{H}/r_{\rm drag}$.  Bottom left: Same but for $D_{V}/r_{\rm drag}$.  Bottom right: Same but for the distance modulus $\mu$. }
    \label{lCDM_distances_hubble}
\end{figure*}

Finally, in Figure \ref{lCDM_distances_hubble}, we display the functions $D_{\rm M}(z_{\rm obs})/r_{\rm drag}$, $D_{\rm H}(z_{\rm obs})/r_{\rm drag}$, $D_{\rm v}(z_{\rm obs})/r_{\rm drag}$, and $\mu(z_{\rm obs})$ (distance modulus) obtained in the standard model using the best-fit DESI-BAO parameters and Planck 2018 parameters \citep{planck2018} alongside those obtained in the CSF model fitting DESI and DESI+CMB+Pantheon+. Notably, when fitting only DESI to the CSF model, the resulting model is able to reproduce the value of $D_{\rm H}(z_{\rm obs}=0.51)$, which deviates from the value predicted by the standard model with Planck 2018 parameters or DESI-Y1 BAO parameters. However, when the CMB and Pantheon+ are included in the fit, the value of $D_{\rm H}(z_{\rm obs}=0.51)$ moves closer to the value predicted by the standard model with Planck or DESI parameters.

Finally, we find that the CSF model obtained from DESI+CMB+Pantheon+ yields results nearly identical to those predicted by the standard model using Planck or DESI parameters for the distance modulus, $\mu(z_{\rm obs})$. Specifically, the ratio between $\mu(z_{\rm obs})$ from the C-$\Lambda$CDM model with parameters constrained from DESI+CMB+Pantheon+ and the standard $\Lambda$CDM model with DESI parameters is approximately 0.25$\%$.

To assess the influence of the value of $D_{H}(z_{\rm obs}=0.51)$ on $\Theta(z_{\rm obs})$, we repeated the analysis by replacing the values of $D_{M}$ and $D_{H}$ from DESI BGS ($z_{\rm obs}=0.30$) with those from SDSS BAO ($z_{\rm obs}=0.38, 0.51, 0.61$) \citep{2017MNRAS.470.2617A}. This test is detailed in Appendix \ref{BOSS}. Figure \ref{theta_BOSS} from that Appendix shows the constrained CSF, $\Theta(z_{\rm obs})$, obtained from DESI-BAO Y1 data and from DESI-BAO+BOSS. Considering the uncertainties, the two constraints are compatible. Additionally, the best-fit value for DESI-BAO+BOSS is more negative than the one obtained with only DESI-BAO, with larger uncertainties.

To conclude this section, we discuss the following theoretical exercise: We imposed that the comoving radial distance $D_{C}(z_{\rm obs})$ obtained in both the CSF model (with known parameters) and the standard model (which parameters we aim to determine) are identical. Our goal was to find the standard model parameters. If the newly determined standard model parameters are sufficiently close to the Planck parameters, we can conclude that the CSF model successfully recovers the Planck parameters.

The methodology to calculate the best-fit parameters and their uncertainties for the standard model is as follows: We imposed that the absolute difference between the comoving radial distance, $D_{C}$ (normalized by $r_{\rm drag}$, given by the CSF model, see equation \ref{DC}), and that given by the standard model (as described in \cite{hogg2000distance}, and also normalized by $r_{\rm drag}$, which has to be determined) is minimized. To do this, we defined a uniform distribution for each CSF parameter ($\Omega_{\rm m}$, H$_{0}$ and $r_{\rm drag}$) centred around its best-fit value and with an amplitude of $\pm$1$\sigma$. We repeated the minimization process 2000 times, each time using a different set of parameter values drawn from these uniform distributions so the value of $D_{C}$(z$_{\rm obs}$) would be different each time. The final best-fit values for $\Omega_{\rm m}$, H$_{0}$, and r$_{\rm drag}$ in the standard model were calculated as the average of the optimal values obtained in these 2000 iterations, and the uncertainties were determined as the standard deviations of these results.

If we consider the CSF $\Lambda$CDM model obtained by fitting DESI+CMB+Pantheon+, the mean values of the parameters from the standard model that minimize the difference of the comoving radial distances of the standard and CSF models are

\begin{equation*}
\left\{
\begin{array}{l}
    \Omega_{\rm m} = 0.323 \pm 0.150 \\
    {\rm H_{0}} = 64.26 \pm 10.03 \\
    r_{\rm drag} = 151.69 \pm 22.21 
\end{array}.
\right.
\end{equation*}

We found that the values of $\Omega_{\rm m}$ and H$_{0}$ are consistent with the Planck 2018 results \citep{planck2018}, although our uncertainties are significantly larger due to the high uncertainties in the CSF parameters.

We then extended this approach to find the parameters for a standard $w$CDM model:

\begin{equation*}
\left\{
\begin{array}{l}
    \Omega_{\rm m} = 0.202 \pm 0.122\\
    {\rm H_{0}} = 71.73 \pm 18.10\\
    r_{\rm drag} = 153.16 \pm 35.84 \\
    w = -1.404 \pm 0.984
\end{array}.
\right.
\end{equation*}

We found that the value of $\Omega_{\rm m}$ is 40$\%$ smaller than that obtained for the $\Lambda$CDM model, while the value of H$_{0}$ is 12$\%$ larger. Once again, the values of $r_{\rm drag}$ obtained for both models are very similar, although the uncertainty for the $w$CDM model is quite large. Moreover, the uncertainty in $w$ remains too large to draw any definitive conclusions about the true value of this parameter for the considered cosmological model.

\begin{figure}[htb!]
    \centering
    \includegraphics[width=0.5\textwidth]{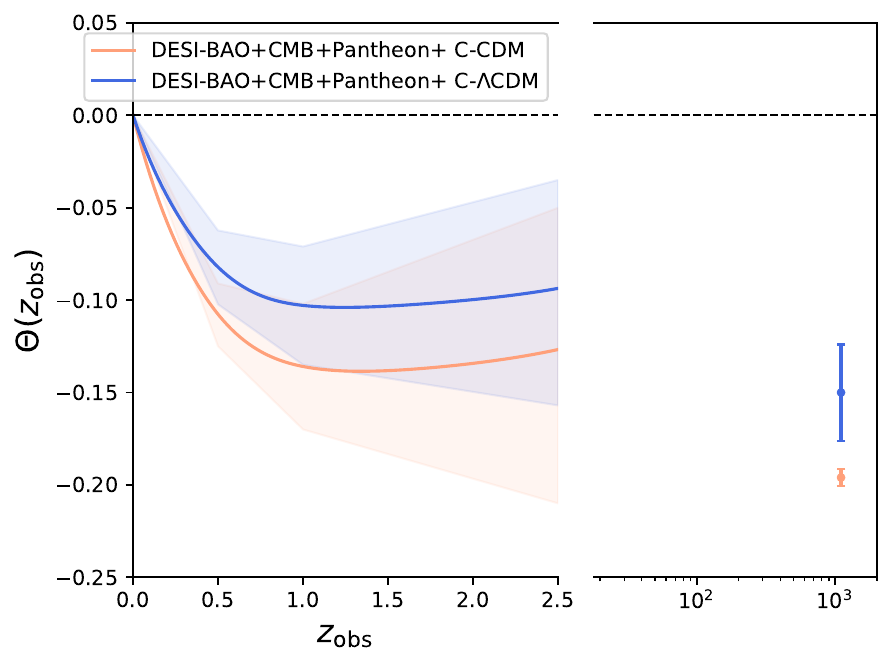}
    \caption{Inferred $\Theta(z_{\rm obs})$ from DESI+CMB+Pantheon+ for a flat C-$\Lambda$CDM model (blue line) and for a flat C-CDM model (orange line). The shadow region delimits the 68$\%$ uncertainty. Both CSF models have been obtained using DESI+CMB+Pantheon+.} 
    \label{theta_CDM}
\end{figure}

Finally, in Appendix \ref{appendixcdm}, we present the constraints on the new parameters and the CSF obtained for a CSF-CDM model (without dark energy). As shown in the aforementioned appendix, the best-fit value of $\Theta$ in the dark-matter-dominated CSF model is systematically more negative than in the C-$\Lambda$CDM model across all redshifts (see Figure \ref{theta_CDM}). This result is particularly relevant, as it reinforces the idea that the CSF encodes information about the missing components of the standard cosmological model. While the values of $\Theta$ remain compatible within 1$\sigma$ for $z_{\rm obs} < 2.5$, the discrepancy in $\Theta_{\rm dec}$ between the two models suggests that the conformal function is sensitive to the absence of dark energy.

To further explore idea that the CSF encodes information about the missing components of the standard cosmological model, we extended our analysis to a CSF-$w_0w_a$CDM model, allowing for a time-dependent dark energy equation of state. However, the large uncertainties in $w_0$ and $w_a$ prevented us from drawing any robust conclusions. In practice, the reconstructed conformal function in this scenario was almost indistinguishable from that obtained in the $\Lambda$CDM case. This highlights a key point. The current model involves too many free parameters when using splines (six parameters, which is excessive), making it difficult to establish a clear relationship between the conformal function and the evolution of $w(a)$.

Finding a direct relationship between the conformal function and $w(a)$ would be highly beneficial, as it would allow us to obtain $w(a)$ directly without making any assumptions about its form, as is typically done when introducing two free parameters, $w_0$ and $w_a$. This would provide a more model-independent approach to understanding the evolution of dark energy.

Moving forward, it is crucial to develop an alternative approach capable of directly probing this relationship in a model-independent manner. Identifying a robust link between the conformal function and the evolution of dark energy remains an open challenge, and we will address this in future work.

One final exercise involved replacing the Pantheon+ sample with DESY5. This is done in Appendix \ref{des_results}, where we show that the uncertainties of all parameters are larger using DESI+CMB+DESY5 than those obtained with DESI+CMB+Pantheon+. This suggests that covering a wider range of redshift is more important than minimizing uncertainties. Pantheon+ has a maximum redshift of $z_{\rm max}$=2.26, while DESY5 has $z_{\rm max}$=1.12. A crucial test to investigate this is repeating the DESI+CMB+Pantheon+ fit while removing all supernovae with $z_{\rm obs}>$1.12 and comparing the results with those from DESI+CMB+DESY5. This exercise is also shown in Appendix \ref{des_results}, where we find that the uncertainties from DESI+CMB+Pantheon+(with $z_{\rm max}$=1.12) are similar to those from DESI+CMB+DESY5. Therefore, this suggests that it is more important to have data across a wide redshift range than to focus on minimizing uncertainties within a narrow range.

\section{Conclusions}
 In this work, we have considered a specific conformal transformation of the FLRW (standard) metric where we averaged the spatial part over large scales to $\langle\Theta(\bar{x})\rangle = 1$ and defined $\Theta(a)$ as the CSF. We also introduced a cubic spline with four nodes and two fixed derivatives as the ratio of the redshift predicted by the FLRW metric to the observed redshift (see equation \ref{zflrw-zobs}), which is directly related to the CSF, $\Theta(z_{\rm obs})$. We constrained these six free parameters along with all the cosmological parameters involved in the BAO distances, supernovae distance modulus, and CMB using DESI-BAO Y1 measurements, the Pantheon+ SN Ia dataset, and the CMB. The key results from this work are as follows:
 
\begin{itemize}
    \item The CSF model is dark matter dominated, with the roles of the dark matter density parameter and the dark energy density parameter exchanged. Specifically, we find $\Omega_{\rm m} \sim 0.7$ and $\Omega_{\Lambda} \sim 0.3$.
    \item The value of H$_{0}$ obtained from DESI+CMB+Pantheon+ in the CSF model, H$_{0} = 64.03 \pm 3.39$, is compatible with the value predicted by Planck 2018, H$_{0} = 67.36 \pm 0.54$, within approximately 1$\sigma$.
    \item The obtained value of $\sigma_{8}$ in the CSF flat $\Lambda$CDM model, $\sigma_{8} = 0.629 \pm 0.029$, is smaller than the Planck 2018 prediction of $\sigma_{8} = 0.8111 \pm 0.0060$ and is only compatible within approximately 6$\sigma$. However, this discrepancy does not imply that the matter power spectrum is not conserved, as a transformation in units must be taken into account. Specifically, the power spectrum must be transformed by multiplying it by $\exp{(-3\Theta_{\rm dec})}$ since its units correspond to distances cubed and by $h_{\rm Planck}/h_{\rm a0}$, where $h_{\rm Planck}$ is the reduced Hubble constant from Planck assuming the standard metric.
    \item The temperature and lensing power spectra obtained with the new cosmological parameters are compatible with observations. However, there is a small difference between the CSF and standard temperature power spectra at low values of $l$, where the CSF model predicts a lower temperature power spectrum than the standard one. A possible interpretation is that the integrated Sachs-Wolfe effect vanishes in the CSF models.
    \item For a CSF-CDM model (without dark energy), the CSF obtained is systematically more negative than in a CSF-$\Lambda$CDM model for all redshifts. This reinforces the idea that the CSF encodes information about the missing components of the cosmological model, as a more negative CSF suggests that more ingredients are absent. Furthermore, we tested a CSF-$w_0w_a$CDM model, but the large uncertainties in $w_0$ and $w_a$ led to a conformal function nearly identical to that of $\Lambda$CDM, thus preventing us from drawing strong conclusions. This highlights the need for an alternative approach to establish a direct dependence between the CSF and $w(a)$ without prior assumptions, which we will explore in future work.
    \item Finally, we attempted to recover the Planck parameters using the CSF model. While we have demonstrated that it is possible to recover Planck’s parameters, the large uncertainties prevent us from drawing any definitive conclusions.
\end{itemize}

Given that the results for the model with and without dark energy are different, we expect to find a way to test the evolution of dark energy, i.e., the dependence of  w(a), using this framework. This will be the objective of a future work.

\begin{acknowledgements}
    EFG acknowledges financial support from the Severo Ochoa grant CEX2021-001131-S funded by MCIN/AEI/ 10.13039/501100011033\\
    JLCC acknowledges support from Conahcyt project CBF2023-2024-589.
    
    EFG and FP acknowledge support from the Spanish MICINN funding grant PGC2018-101931-B-I00. 

    We are grateful to Eric Linder for taking the time to read our paper and provide valuable feedback. We especially appreciate his assistance in helping us writing the equations describing the evolution of density with the scale factor in the CSF model.

    We would like to express our sincere gratitude to Tamara M. Davis for her valuable contributions to this work. Her assistance with the analysis of the DESY5 SN Ia data was instrumental, and her insightful comments on the first draft significantly improved the clarity and quality of the manuscript.
    
    The authors are honoured to be permitted to conduct scientific research on Iolkam Du’ag (Kitt Peak), a mountain with particular significance to the Tohono O’odham Nation.
\end{acknowledgements}


%
   \bibliographystyle{aa} 
   \bibliography{example} 
%

\begin{appendix}

\section{Best-fit values obtained with DESY5 SN Ia a CSF $\Lambda$CDM model}\label{des_results}
In this section, we present the results obtained by fitting the CSF $\Lambda$CDM model using DESI-Y1 BAO measurements, CMB, and DESY5 SN Ia, instead of Pantheon+. The comparison between DESI+CMB+Pantheon+ and DESI+CMB+DESY5 is shown in the second and fourth columns of Table \ref{des}, respectively. It can be seen that the best-fit value of $\Omega_{\rm m}$ obtained with DESI+CMB+Pantheon+ is smaller than that obtained with DESI+CMB+DESY5. However, both values are compatible within 1$\sigma$ when considering the uncertainties. This is consistent with the fact that Pantheon+ predicts a smaller value for $\Omega_{\rm m}$ (0.334 $\pm$ 0.018) \citep{2022ApJ...938..110B}, compared to DESY5 (0.352 $\pm$ 0.017) \citep{2024ApJ...973L..14A} for the standard $\Lambda$CDM model.

We also observe that the uncertainties for the CSF parameters $\Theta(z_{\rm obs}=0.5, 1.0, 2.5)$ are larger when using DESI+CMB+DESY5 than when using DESI+CMB+Pantheon+. The same trend is observed for $\Omega_{\rm m}$ and H$_{0}$. For the other parameters, the uncertainties are similar. This can be explained by the fact that DESY5 covers a smaller redshift range ($z{\rm max}=1.12$) compared to Pantheon+ ($z_{\rm max}=2.26$), although the uncertainties in $\mu(z_{\rm obs})$ from DESY5 are smaller than those from Pantheon+.

\begin{table*}[h!]
    \caption{Posterior mean and standard deviation for the parameters of the CSF flat $\Lambda$CDM models from DESI-BAO Y1+CMB+Pantheon+ and DESI+CMB+DESY5.}
    \centering
    \renewcommand{\arraystretch}{1.25} 
    \begin{tabular}{c|ccc}
    \toprule
         Parameter &DESI+CMB+Pantheon+ &DESI+CMB+Pantheon+(z$_{\rm max}=1.12$)&DESI+CMB+DESY5\\
         \midrule
         $100\Omega_{\rm b}h^{2}$ &2.631$\pm$0.083 &2.590$\pm$0.074&2.609$\pm$0.076\\
 $\Omega_{\rm c}h^{2}$ &0.137$\pm$0.003 &0.137$\pm$0.004&0.138$\pm$0.004\\
         $\tau$ &0.0508$\pm$0.0069 &0.0514$\pm$0.0074&0.0511$\pm$0.0078\\
         ln($10^{10}A_{\rm s}$) &3.036$\pm$0.014 &3.039$\pm$0.014&3.039$\pm$0.015\\
         $n_{s}$ &0.9708$\pm$0.0063 &0.9680$\pm$0.0043&0.9681$\pm$0.0042\\
         $\Theta_{\rm dec}$ &-0.152$\pm$0.025 &-0.141$\pm$0.027&-0.149$\pm$0.027\\
         $\Theta(0.5)$ &-0.084$\pm$0.019 &-0.078$\pm$0.019&-0.080$\pm$0.020\\
         $\Theta(1.0)$ &-0.106$\pm$0.032 &-0.097$\pm$0.031&-0.099$\pm$0.032\\
         $\Theta(2.5)$ &-0.099$\pm$0.061 &-0.088$\pm$0.061&-0.089$\pm$0.064\\
         $d\Theta/dz_{\rm obs}$(0.0) &-0.281$\pm$0.054 &-0.263$\pm$0.056&-0.266$\pm$0.057\\
         $d\Theta/dz_{\rm obs}$(2.5) &0.014$\pm$0.020 &0.016$\pm$0.021&0.020$\pm$0.022\\
         H$_{a0}$ &55.30$\pm$1.98 &48.39$\pm$3.38&47.45$\pm$3.37\\
         \midrule
         H$_{0}$ &63.90$\pm$3.34 &64.36$\pm$3.66&61.91$\pm$4.42\\
         $\Omega_{\rm m}$ &0.733$\pm$0.099 &0.711$\pm$0.120&0.745$\pm$0.124\\
         $h_{a}r_{d}$ &72.60$\pm$5.12 &74.05$\pm$5.60&72.97$\pm$5.60\\
         $\sigma_{8}$ &0.637$\pm$0.031 &0.651$\pm$0.032&0.642$\pm$0.032\\
        $r_{d}/$Mpc &152.60$\pm$5.12 &153.07$\pm$5.98&153.84$\pm$6.44\\
        Age [Gyr]  &13.08$\pm$0.14 &13.15$\pm$0.12&13.12$\pm$0.12\\
        \bottomrule
        \end{tabular}
    \tablefoot{Posterior mean and standard deviation for the parameters of CSF flat $\Lambda$CDM models. The first column shows the constraints from using DESI-BAO Y1 combined with CMB and Pantheon+, and the second one shows the constraints  obtained combining DESI+CMB+DESY5 in the CSF model.}
    \label{des}
\end{table*}

An exercise to assess the constraining power of high-redshift supernovae from Pantheon+ is to consider only those with $z_{\rm obs} < 1.12$ and repeat the fits using this new sample. This would allow us to determine whether it is more important to have data spanning a wide redshift range, or to focus on a narrower range with smaller uncertainties.

The best-fit values obtained from DESI+CMB+Pantheon+, but considering only those supernovae with $z_{\rm obs} < 1.12$, are shown in the third column of Table \ref{des}. From this table, we can see that the uncertainties in all parameters are very similar to those obtained with DESI+CMB+DESY5. This indicates that having data across a wide redshift range provides more constraining power than reducing the uncertainties in the magnitude measurements.

\begin{figure}[htb!]
    \centering
    \includegraphics[width=0.5\textwidth]{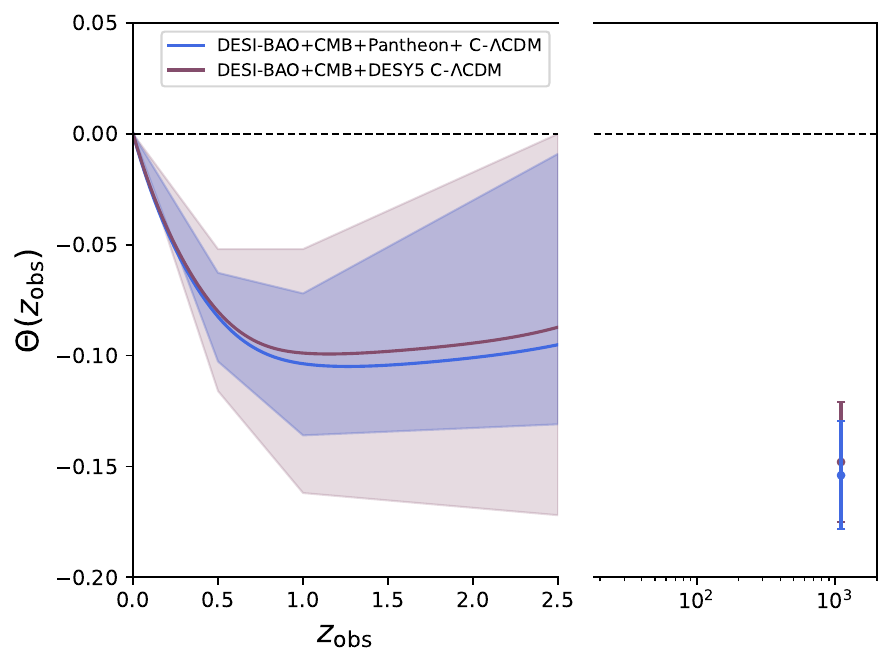}
    \caption{$\Theta(z_{\rm obs})$ inferred from DESI+CMB+Pantheon+ and DESI+CMB+DESY5 for a flat C-$\Lambda$CDM model. The shadow region delimits the 68$\%$ uncertainty.} 
    \label{theta_DES}
\end{figure}

\begin{figure*}[htb!]
    \centering
    \includegraphics[width=0.495\textwidth]{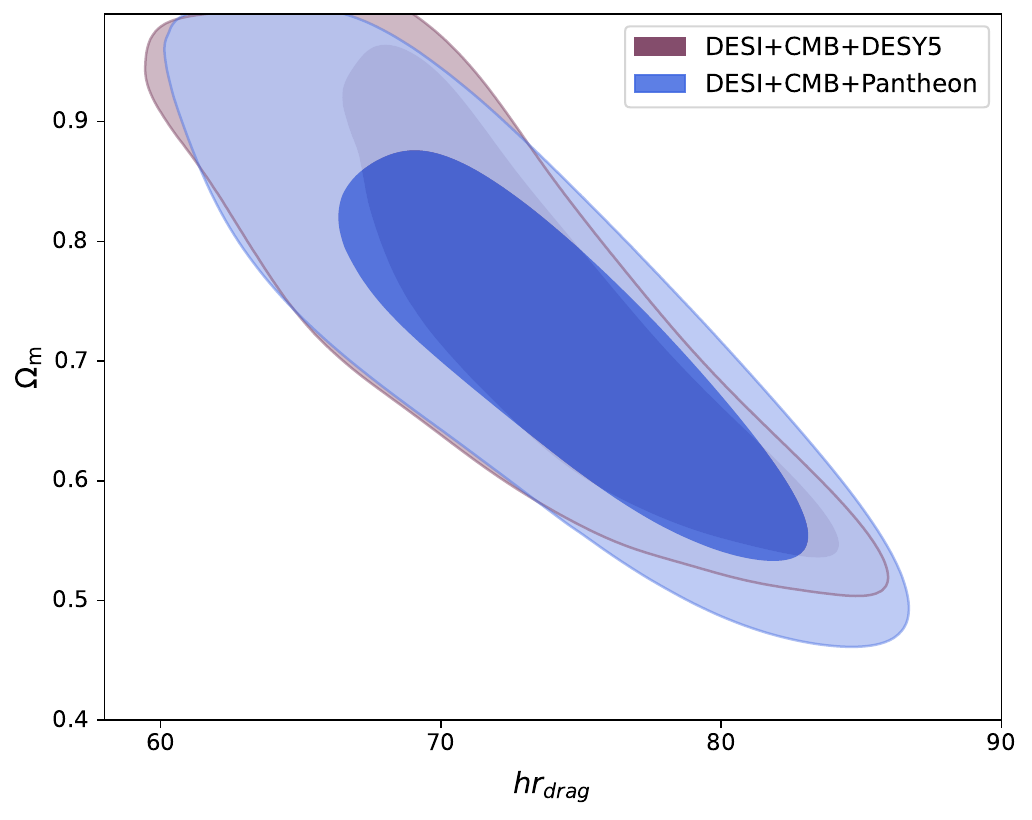}
    \includegraphics[width=0.49\textwidth]{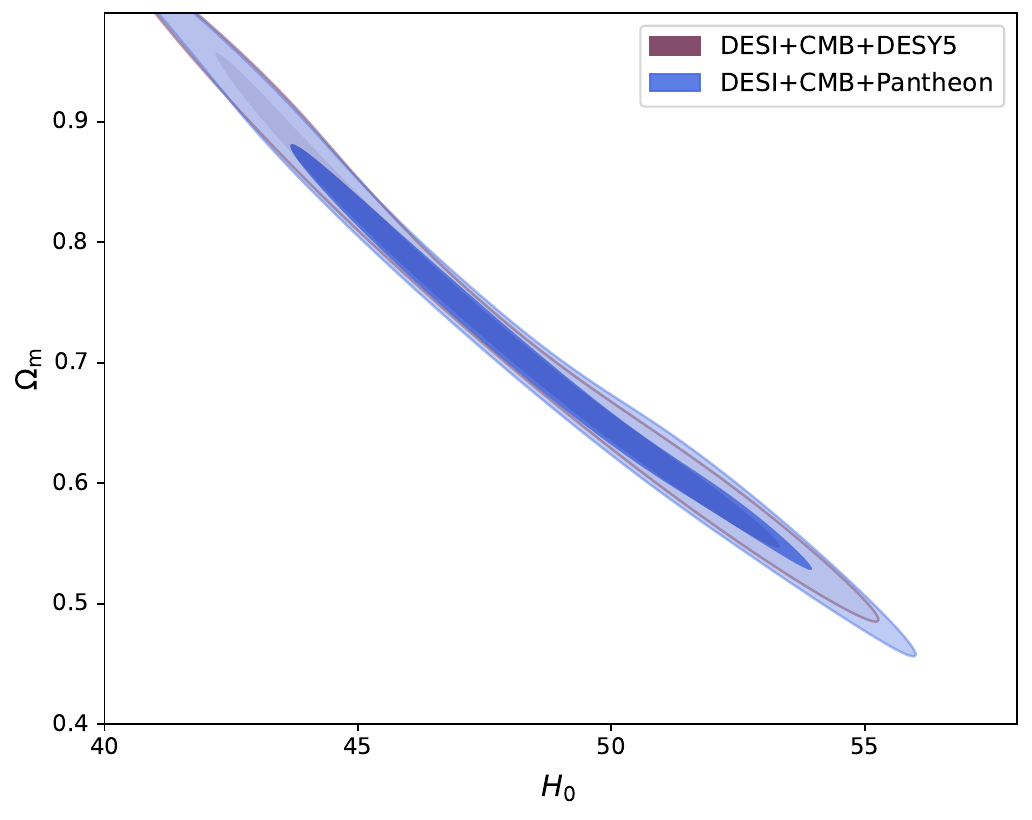}
    \caption{Left: Marginalized 68\% and 95\% confidence level constraints on the matter density parameter, $\Omega_{\rm m}$ and the Hubble constant H$_{0}$ multiplied by the sound horizon scale $r_{\rm drag}$ assuming a CSF flat $\Lambda$CDM model using DESI+CMB+Pantheon+ and DESI+CMB+DESY5 Right: the same as in the left panel but for the plane $\Omega_{\rm m}-$H$_{a0}$ plane.}
    \label{des_contour}
\end{figure*}

\section{Flat CSF dark-matter-dominated models (C-CDM)}\label{appendixcdm}
In this Appendix, we consider a dark-matter-dominated universe with a CSF transformation (C-CDM) and examine how all the cosmological parameters change, especially $\Theta(z_{\rm obs})$. If $\Theta(z_{\rm obs})$ deviates from 0 (i.e., if it becomes more negative than in the C-$\Lambda$CDM model), this would suggest that $\Theta(z_{\rm obs})$ carries information about dark energy and, more broadly, about the missing components of the cosmological model.

Figure \ref{theta_CDM} shows the CSF obtained for the C-CDM model, alongside the result for the C-$\Lambda$CDM model. In both cases, we have used DESI+CMB+Pantheon+ data to derive $\Theta(z_{\rm obs})$ and $\Theta_{\rm dec}$. The best-fit value of $\Theta$ in the dark-matter-dominated CSF model is more negative than that in the C-$\Lambda$CDM model for all redshifts. However, accounting for uncertainties, the values from both models are compatible within 1$\sigma$ for $z_{\rm obs} < 2.5$. A closer look at $\Theta_{\rm dec}$ reveals that the values from both models are not compatible within 1$\sigma$. Table \ref{values_appendix} presents the best-fit values obtained by fitting only DESI-BAO and only the CMB to the CSF-CDM model.

Indeed, we observe that the CSF model can provide insights into the missing components of the standard model. Specifically, if we exclude dark energy from our CSF model, we obtain a different, more negative CSF compared to the one derived from a model that includes dark energy. This is crucial, as it indicates that the CSF is capable of revealing the missing elements of the cosmological model in question, and it does so in the expected manner. If a cosmological model included all the components that are currently absent in the standard model, we would expect to see a constant zero CSF. However, identifying these missing components falls outside the scope of our current work.

We can replicate the exercise of calculating the parameters for the standard metric model by using the CSF-CDM (no dark energy) model fitted to the DESI+CMB data to recover the Planck parameters. The parameters we recover for a standard flat $\Lambda$CDM model are

\begin{equation*}
\left\{ 
\begin{array}{l}
    \Omega_{\rm m} = 0.330 \pm 0.179\\
    {\rm H_{0}} = 68.54 \pm 10.37\\
    r_{\rm drag} = 150.71 \pm 20.78
\end{array}. 
\right.
\end{equation*}

The obtained values for $\Omega_{\rm m}$, H$_{0}$, and $r{\rm drag}$ are consistent with the Planck 2018 results \citep{planck2018} within 1$\sigma$.

Next, we delve deeper and determine the parameters for a standard $w$CDM model:

\begin{equation*}
\left\{
\begin{array}{l}
    \Omega_{\rm m} = 0.215 \pm 0.149 \\
    {\rm H_{0}} = 75.32\pm 17.25 \\
    r_{\rm drag} = 152.08 \pm 33.81 \\    
    w = -1.42 \pm 1.02
\end{array}.
\right.
\end{equation*}

We observe that the value of $\Omega_{\rm m}$ is approximately 35$\%$ smaller than that obtained for the CDM model, while the value of H$_{0}$ is about 10$\%$ larger than in the CDM model. The values of $r{\rm drag}$ are nearly identical for both models, although the uncertainties in the $w$CDM model for this parameter are significantly larger.

\begin{table*}[h!]
\caption{Posterior mean and standard deviation for the parameters of the flat CSF C-CDM model from DESI-BAO Y1, CMB with CSF, and DESI+CMB+Pantheon+.}
    \centering
    \renewcommand{\arraystretch}{1.25} 
    \begin{tabular}{c|ccc}
    \toprule
         Parameter&  DESI  &CMB &DESI+CMB+Pantheon+\\
         \midrule
         $100\Omega_{\rm b}h^{2}$&  -- &2.7414$\pm$0.0003&2.742$\pm$0.027\\
         $\Omega_{\rm c}h^{2}$&  -- &0.1439$\pm$0.0008&0.1439$\pm$0.0009\\
         $\tau$&  -- &0.0496$\pm$0.0076&0.0498$\pm$0.0077\\
         ln($10^{10}A_{\rm s}$)&  -- &3.035$\pm$0.014&3.036$\pm$0.015\\
         $n_{s}$&  -- &0.9691$\pm$0.0041&0.9693$\pm$0.0042\\
         $\Theta_{\rm dec}$&  -- &-0.197$\pm$0.004&-0.197$\pm$0.005\\
         $\Theta(0.5)$& -0.099$\pm$0.017&-- &-0.108$\pm$0.017\\
         $\Theta(1.0)$& -0.125$\pm$0.034&-- &-0.136$\pm$0.033\\
         $\Theta(2.5)$& -0.107$\pm$0.080&-- &-0.129$\pm$0.078\\
         $d\Theta/dz_{\rm obs}$(0.0)& -0.307$\pm$0.050&-- &-0.363$\pm$0.034\\
         $d\Theta/dz_{\rm obs}$(2.5)& 0.030$\pm$0.027&-- &-0.017$\pm$0.026\\
         H$_{a0}$& -- &41.50$\pm$0.08&41.49$\pm$0.08\\
         \midrule
         H$_{0}$& -- & -- &59.92$\pm$2.22\\
         $h_{a}r_{d}$& 65.01$\pm$3.36&-- &63.97$\pm$3.27\\
        $\sigma_{8}$& -- &0.583$\pm$0.006&0.583$\pm$0.006\\
 $r_{d}/$Mpc& -- & 147.38$\pm$0.26&154.15$\pm$7.89\\
 Age [Gyr] & -- & 12.90$\pm$0.04&12.90$\pm$0.04\\
         \bottomrule

        \end{tabular}
    \tablefoot{Posterior mean and standard deviation for the parameters of flat CSF CDM C-CDM model. The first column shows the constraints from using only DESI-BAO Y1 measurements, the second one shows the constraints for only CMB with CSF and the third one shows the constraints obtained combining DESI+CMB+Pantheon+ in the CSF model.}
    \label{values_appendix}
\end{table*}

\section{Comparison of DESI results with DESI+SDSS BAO data}\label{BOSS}
In this Appendix, we present the results obtained by substituting the values of $D_{M}$ and $D_{H}$ from DESI BGS ($z_{\rm obs}=0.30$) with those from SDSS BAO ($z_{\rm obs}=0.38, 0.51, 0.61$) \citep{2017MNRAS.470.2617A}. This test is significant, as Figure \ref{lCDM_distances_hubble} reveals a substantial discrepancy between the value of $D_{H}/r_{\rm drag}$ measured from DESI Y1 and that predicted by a $\Lambda$CDM model with Planck 2018 parameters \citep{planck2018}.

\begin{figure}[htb!]
    \centering
    \includegraphics[width=0.5\textwidth]{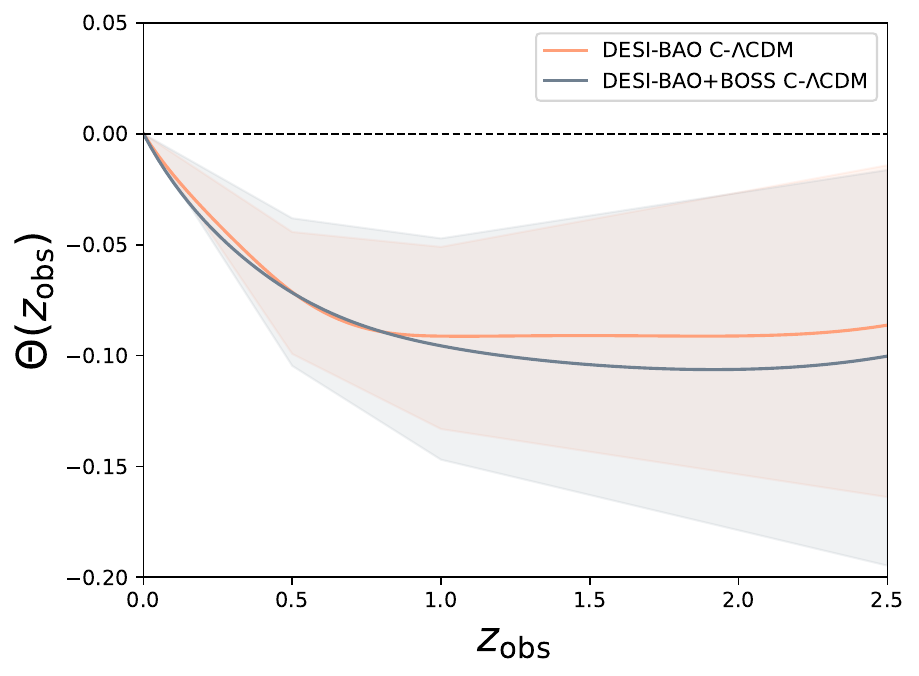}
    \caption{$\Theta(z_{\rm obs})$ inferred from DESI-BAO Y1 distance measurements and DESI-BAO+BOSS for a flat C-$\Lambda$CDM model and for a flat C-$\Lambda$CDM model (black line). The shadow region delimits the 68$\%$ uncertainty.} 
    \label{theta_BOSS}
\end{figure}

Figure \ref{theta_BOSS} shows the constrained CSF, $\Theta$, obtained with DESI-BAO Y1 data and DESI-BAO+BOSS. Taking into account the uncertainties, both constraints are compatible. Furthermore, the best-fit value for DESI-BAO+BOSS is more negative than that obtained with DESI-BAO alone, although with larger uncertainties.

Table \ref{boss_values} presents the best-fit values obtained for each case. The uncertainties for DESI-BAO+BOSS are larger than those for DESI-BAO alone. Thus, the two sets of constrained parameters are compatible within 1$\sigma$.

\begin{table}[htb!]
    \caption{Comparison of the posterior mean and standard deviation for the flat CSF $\Lambda$CDM model using DESI-BAO Y1 and DESI-BAO+BOSS.}
    \centering
    \renewcommand{\arraystretch}{1.25} 
    \begin{tabular}{c|cc}
    \toprule
         Parameter&  DESI  &DESI+BOSS\\
         \midrule
         $\Theta(0.5)$& -0.070$\pm$0.027 &-0.072$\pm$0.034\\
         $\Theta(1.0)$& -0.090$\pm$0.041 &-0.097$\pm$0.050\\
         $\Theta(2.5)$& -0.088$\pm$0.074 &-0.104$\pm$0.089\\
         $d\Theta/dz_{\rm obs}$(0.0)& -0.206$\pm$0.089 &-0.253$\pm$0.104\\
         $d\Theta/dz_{\rm obs}$(2.5)& 0.0204$\pm$0.025 &0.022$\pm$0.026\\
         $\Omega_{\rm m}$& 0.685$\pm$0.183 &0.622$\pm$0.211\\
         $\Omega_{\Lambda}=1-\Omega_{\rm m}$& 0.315$\pm$0.183 &0.378$\pm$0.211\\
         $h_{a}r_{d}$& 75.92$\pm$8.84 &76.83$\pm$10.89\\
         \bottomrule
        \end{tabular}
    \tablefoot{Comparison of the posterior mean and standard deviation obtained for the parameters of a flat CSF $\Lambda$CDM model using only DESI-BAO Y1 measurements (first column) and DESI-BAO+BOSS (second column).}
    \label{boss_values}
\end{table}

In conclusion, although there is a significant discrepancy between the BGS BAO distance obtained with DESI Y1 and the $\Lambda$CDM model, this discrepancy is not observed in the BOSS data. However, the large uncertainties in the BOSS data prevent definitive conclusions.

\end{appendix}

\end{document}